%% file: paper.tex
\documentclass[conference,10pt]{IEEEtran}
\IEEEoverridecommandlockouts
\newcommand{\forarxivversion}[1]{#1}  	% set this to {} to hide the references to appendices 
\newcommand{\forlicsversion}[1]{} %{#1} 			% set this to {#1} to use up the little bit of space you get by deleting references to the appendices
\newcommand{\todo}[1]{} %{\textcolor{red}{[ToDo: #1]}}

\usepackage{graphicx} % Required for inserting images
\usepackage{amsfonts}
\usepackage{amsthm}
\usepackage{amsmath}
\usepackage{amssymb}
\usepackage{amscd}
\usepackage{mathtools}
\usepackage{stmaryrd}
\usepackage{upgreek}
\usepackage{mathabx} % for \Asterisk
\usepackage{mathpartir}
\usepackage{multirow}
\usepackage{quiver}
\usepackage{csquotes} % for quote blocks
\usepackage{framed}
\usepackage[inline]{enumitem} % for inlining enumerates
\usepackage{subcaption}
\usepackage{bussproofs}
\usepackage{accents}
\newcommand*{\dt}[1]{%
  \accentset{\mbox{\large\bfseries .}}{#1}}

\input{phaa.sty}

\usepackage{hyperref}
\usepackage{cleveref}[capitalize]
\crefname{Th}{theorem}{theorems}
\Crefname{Th}{Theorem}{Theorems}
\crefname{Lem}{lemma}{lemmas}
\Crefname{Lem}{Lemma}{Lemmas}
\crefname{Cor}{corollary}{corollaries}
\Crefname{Cor}{Corollary}{Corollaries}
\crefname{Prop}{proposition}{propositions}
\Crefname{Prop}{Proposition}{Propositions}
\crefname{Def}{definition}{definitions}
\Crefname{Def}{Definition}{Definitions}
\crefname{Rem}{remark}{remarks}
\Crefname{Rem}{Remark}{Remarks}
\crefname{Ex}{example}{examples}
\Crefname{Ex}{Example}{Examples}
\crefname{figure}{figure}{figures}
\Crefname{figure}{Figure}{Figures}
\crefname{construction}{construction}{constructions}
\Crefname{construction}{Construction}{Construction}
\crefname{appendix}{appendix}{appendices}
\Crefname{appendix}{Appendix}{Appendices}
\usepackage{thm-restate}

\newtheorem{theorem}{Theorem}[section]
\newtheorem{lemma}[theorem]{Lemma}
\newtheorem{corollary}[theorem]{Corollary}
\newtheorem{proposition}[theorem]{Proposition}

\theoremstyle{definition}
\newtheorem{definition}[theorem]{Definition}

\newtheorem{remark}[theorem]{Remark}
\newtheorem{?}[theorem]{Problem}
\newtheorem{example}[theorem]{Example}
\newtheorem{notation}[theorem]{Notation}
\newtheorem{construction}[theorem]{Construction}

\newcommand{\mypara}[1]{\vspace{2mm}{\it #1}}

\renewcommand{\epsilon}{\varepsilon}
\renewcommand{\subset}{\subseteq}

\newcommand{\eg}{e.g.}
\newcommand{\ie}{i.e.}
\newcommand{\cf}{cf.}
\newcommand{\ibid}{ibid.}

\input{macros}

\newenvironment{bprooftree} % put prooftrees in a box for side by side
  {\leavevmode\hbox\bgroup}
  {\DisplayProof\egroup}

\begin{document}

\title{Logical relations for call-by-push-value models, \\ via internal fibrations in a 2-category
	\thanks{PAdA was supported by the ERC Consolidator Grant BLAST, and the ARIA programme on Safeguarded AI.  SK was supported by JST ACT-X Grant Number JPMJAX2104; JSPS Overseas Research Fellowships. PS was partially supported by the Air Force Office of Scientific Research
						(award number FA9550-21-1-0038).}
}

\author{\IEEEauthorblockN{Pedro H. Azevedo de Amorim}
\IEEEauthorblockA{\textit{Dept. of Computer Science} \\
\textit{University of Oxford}\\
Oxford, UK \\
pedro.azevedodeamorim@cs.ox.ac.uk}
\forarxivversion{\vspace{-5mm}}
\and
\IEEEauthorblockN{Satoshi Kura}
\IEEEauthorblockA{\textit{Fac.\ of Ed.\ and Integrated A\&S} \\
\textit{Waseda University}\\
Tokyo, Japan \\
satoshi.kura@aoni.waseda.jp}
\forarxivversion{\vspace{-5mm}}
\and
\IEEEauthorblockN{Philip Saville}
\IEEEauthorblockA{\textit{Dept. of Informatics} \\
\textit{University of Sussex}\\
Brighton, UK \\
p.saville@sussex.ac.uk}
\forarxivversion{\vspace{-5mm}}
} 

\maketitle

\begin{abstract}
We give a denotational account of logical relations for call-by-push-value (CBPV) in the fibrational style of Hermida, Jacobs, Katsumata and others. Fibrations—which axiomatise the usual notion of sets-with-relations—provide a clean framework for constructing new, logical relations-style, models. Such models can then be used to study properties such as effect simulation.

Extending this picture to CBPV is challenging: the models incorporate both adjunctions and enrichment, making the appropriate notion of fibration unclear. We handle this using 2-category theory. We identify an appropriate 2-category, and define CBPV fibrations to be fibrations internal to this 2-category which strictly preserve the CBPV semantics.

Next, we develop the theory so it parallels the classical setting. We give versions of the codomain and subobject fibrations, and show that new models can be constructed from old ones by pullback. The resulting framework enables the construction of new, logical relations-style, models for CBPV.

Finally, we demonstrate the utility of our approach with particular examples. These include a generalisation of Katsumata’s $\toptop$-lifting to CBPV models, an effect simulation result, and a relative full completeness result for CBPV without sum~types. 
\end{abstract}

\section{Introduction}

This paper is about extending the denotational theory of logical relations from effectful call-by-value languages to Levy's call-by-push-value \cite{levy2001thesis,levy2003book}. Logical relations are a fundamental tool for proving metatheoretic properties of logics and programming languages. We begin with a brief overview; a more detailed account can be found in \eg\ \cite[\S2.2]{hermida:thesis}.  

In its simplest form, a logical relation $\oR$ for a typed programming language consists of a predicate $\oR_\oA \subset \sem{\oA}$ on the (set-theoretic) interpretation of each type $\oA$, such that the predicate at complex types is determined inductively by a \emph{logical relations condition}. This condition typically encodes the elimination rules of the corresponding type. This is particularly clear in operationally-motivated examples, where $\sem{\oA}$ is a set of closed terms of type $\oA$. For product types 
	$\oA_1 \times \oA_2$, 
for example, one typically requires that 
	$\oR_{\oA_1 \times \oA_2}$
consists of those $\tM$ such that the projections 
	$\pi_i(\tM)$ 
are elements of  $\oR_{\oA_i}$ for $i = 1, 2$.
The key property of logical relations---which follows from the logical relations conditions---is that they are determined by their base types: if the interpretation $\sem{\oc} \in \sem{\beta}$ of every constant
	$\diamond \vdash \oc : \beta$ 
is in the relation---so that $\sem{\tM} \in \oR_\beta$---then for every closed term $\tN$ we have 
	$\sem{\tN} \in \oR_\oA$.
This fact, which is often proven by induction on the terms, is called the \emph{basic lemma} of logical relations. 

Viewed in this way, logical relations are already a powerful tool. For example, Sieber used a similar construction to show that Plotkin's parallel-or function~\cite{plotkin:LCF-considered-as-a-programming-lang} is not definable in the standard domains model of PCF~\cite{sieber}.
Moreover, a host of sophisticated refinements have extended these ideas to reason about subtle properties of rich languages (\eg\ \cite{appel2001:step-indexed-LRs,pitts2010:tutorial,dreyer2011:logical-step-indexed-LRs}). 

Logical relations also fit into a very general denotational story, originally due to Ma~\&~Reynolds \cite{ma-reynolds:types-abstraction-and-parametric-polymorphism} and Mitchell \& Scedrov~\cite{mitchell-scedrov:notes-on-sconing-and-relators} and extended by many authors since (\eg~\cite{jung-tiuryn:a-new-characterization-of-lambda-definability,hermida:thesis,alimohamed:a-characterisaton-of-lambda-definability,jacobs:categorical-logic-and-type-theory,katsumata:a-characterisation-of-lambda-def-with-sums}). 
The central technical aim of this paper is to extend this story to call-by-push-value (CBPV)~\cite{levy2001thesis,levy2003book}.
Achieving this requires two main technical steps. First,  defining an appropriate notion of `fibration for logical relations' (\cf~\cite{katsumata:relating-computational-effects-by-TT-lifting}) by restricting to fibrations which strictly preserve the model structure. Second, showing that we can universally construct new fibrations for logical relations from old ones. 
The first step tells you how to define logical relations; the second shows how to construct a wide variety of examples.

To explain these steps, and the obstructions to extending this straightforwardly to CBPV, we begin by outlining the story in two simpler cases, namely the simply-typed $\uplambda$-calculus (STLC) and Moggi's monadic metalanguage \monML~\cite{moggi-monads}. We assume some familiarity with (Grothendieck) fibrations and their theory; for a detailed introduction see~\cite{jacobs:categorical-logic-and-type-theory,hermida:thesis}.

\subsection{Logical relations from fibrations for logical relations}

First let us sketch how relations models---and, more generally, fibrations for logical relations---determine logical relations. 
The high-level picture is as follows. One starts with a (category-theoretic) model $\catM$ and constructs a \emph{relations model} $\lift\catM$, together with a functor $\fp : \lift\catM \to \catM$ which strictly preserves the model structure. The objects of $\catM$ are thought of as objects paired with relations, and the morphisms as maps preserving those relations. Following Jacobs~\cite{jacobs:categorical-logic-and-type-theory} and Hermida~\cite{hermida:thesis}, we encode this relation-like structure by asking for $\fp$ to be a fibration;
we then say a \emph{fibration for logical relations} is a fibration which strictly preserves the model structure.

Within this framework the logical relations conditions are embodied by the model structure of $\lift\catM$, the relations $\oR_\oA$ are replaced by an interpretation $\sem{\oA}^{\lift\catM}$ in $\lift\catM$, and the basic lemma may be proven by induction---or, more abstractly, follows from the initiality of the syntactic model. Indeed, choosing the relation $\oR_\beta$ for each base type $\beta$ above amounted to choosing an interpretation of the base types in a relational model. So long as the interpretation $\sem{\beta}^{\lift\catM}$ of base types and constants in $\lift\catM$ lie above their  interpretation $\sem{\beta}^{\catM}$ in $\catM$---in the sense that 
	$\sem{\beta}^{\catM} = \fp( \sem{\beta}^{\lift\catM} )$---then, because $\fp$ strictly preserves the model structure, we get $\sem{\oA}^{\catM} = \fp( \sem{\oA}^{\lift\catM} )$ for every type $\oA$.

\begin{example}
	\label{ex:logical-relations-for-stlc}
	Consider STLC with a single base type $\beta$, together with its usual interpretation in the cartesian closed category $\Set$ of sets and functions (see \eg~\cite{crole1994catsfortypes}). 
	A natural choice of relations model is the category $\Pred$, which has objects pairs $(\oX, \overline\oX)$ consisting of a set $\oX$ and a predicate $\overline\oX \subset \oX$, and morphisms 
		$(\oX, \overline\oX) \to (\oY, \overline\oY)$
	given by functions $\mf : \oX \to \oY$ preserving the relation:
		$x \in \overline\oX \implies \mf(x) \in \overline\oY$.
	This category is cartesian closed, and the forgetful functor $\fp : \Pred \to \Set$ strictly preserves this structure (\eg\ \cite{katsumata:relating-computational-effects-by-TT-lifting}).
	
	Now choose a predicate $\oR_\beta \subset \sem{\beta}$ on the interpretation of $\beta$. This amounts to choosing $(\sem{\beta}, \oR_\beta) \in \Pred$ such that 
		$\fp(\sem{\beta}, \oR_\beta) = \sem{\beta}$.
	Since $\fp$ strictly preserves cartesian closed structure, 
		$\fp(\sem{\tM}^{\Pred}) = \sem{\tM}^{\Set}$
	for every term $\tM$. Setting
		$\sem{\oA}^{\Pred} := (\sem{\oA}^{\Set}, \oR_\oA)$,
	the family $\{ \oR_\oA \mid \oA \text{ a type} \}$ is exactly a logical relation in the sense sketched above.
\end{example}

In recent years there has been extensive work  extending fibrational techniques to Moggi-style monadic models of call-by-value (CBV) languages (\eg~\cite{goubault-larrecq:logical-relations-for-monadic-types-csl,goubault-larrecq:logical-relations-for-monadic-types,katsumata:semantic-formulation-of-TT-lifting,kammar-mcdermott:factorisation-systems-for-logical-relations,katsumata2018codensity,kks:popl2022-full-abstraction}). 
The next example is a simple instance of this framework.

\begin{example}
	\label{ex:logical-relations-for-monadic-metalang}
	We extend \Cref{ex:logical-relations-for-stlc} from STLC to \monML. A semantic model is now a cartesian closed category equipped with a \emph{strong monad}~\cite{kock:strong-functors-and-monoidal-monads} (for an overview, see~\cite{moggi-monads,mcdermott2022whatmakes}).  
	We take the exception monad $\Exc := (-) + \oE$ on $\Set$, where $\oE$ is a fixed set of exception names. Since $\Pred$ has coproducts, which we denote
		$\smallsum_{i=1}^n (\oX_i, \overline{\oX}_i) 
					 := \big( \smallsum_{i=1}^n \oX_i, \bigoplus_{i=1}^n \overline{\oX_i} \big)$,
	for any subset $\overline\oE \subset \oE$ we get a (strong) exception monad 
			$\lift\Exc := (-) + (\oE, \overline\oE)$ 
	which is strictly preserved by $\fp : \Pred \to \Set$. 
	Hence $\fp$ preserves the interpretation of every monadic metalanguage term and, as above, we may define a logical relation: writing 
		$\sem{\oA}^{\Pred} := (\sem{\oA}, \oR_{\oA})$ 
	for all types $\oA$, we now get  
		$\oR_{\monT\oA} := \oR_{\oA} \oplus \overline{\oE}$.
\end{example}

\subsection{Constructing fibrations for logical relations}
\label{sec:constructing-fib-for-log-rel}

We have seen that fibrations for logical relations encode logical relations. But how do we construct them in practice?
There is a canonical way to do this.
Building on \Cref{ex:logical-relations-for-stlc}, say a fibration $\fp : \catE \to \catB$ is an \emph{STLC fibration} if both $\catE$ and $\catB$ are cartesian closed and $\fp$ strictly preserves the cartesian closed structure: this is the appropriate form of fibrations for logical relations for STLC.
One then observes the following.  
 
\begin{lemma}[{\eg~\cite[\S4.3.1]{hermida:thesis}}]
	\label{res:STLC-fibrations-closed-under-pullback}
	Let $\fp : \catE \to \catB$ be an STLC fibration and $\funF : \catC \to \catB$ be any product-preserving functor. Then the pullback of $\fp$ along $\funF$ is also an STLC fibration.
%	\[
%	% https://q.uiver.app/#q=WzAsNCxbMCwxLCJcXGNhdEEiXSxbMSwxLCJcXGNhdEIiXSxbMSwwLCJcXGNhdEUiXSxbMCwwLCJcXG1hdGhiYntQfSJdLFswLDEsIlxcZnVuRiIsMl0sWzIsMSwiXFxmcCJdLFszLDAsIlxcZnEiLDJdLFszLDJdLFszLDEsIiIsMSx7InN0eWxlIjp7Im5hbWUiOiJjb3JuZXIifX1dXQ==
%	\begin{tikzcd}
%		{\mathbb{P}} & \catE \\
%		\catC & \catB
%		\arrow[from=1-1, to=1-2]
%		\arrow["\fq"', from=1-1, to=2-1]
%		\arrow["\lrcorner"{anchor=center, pos=0.125}, draw=none, from=1-1, to=2-2]
%		\arrow["\fp", from=1-2, to=2-2]
%		\arrow["\funF"', from=2-1, to=2-2]
%	\end{tikzcd}
%	\]
\end{lemma}

In various levels of generality, this has been called \emph{sconing}~(\eg~\cite{mitchell-scedrov:notes-on-sconing-and-relators,freyd2006:categories-allegories,hermida:thesis}), \emph{(Artin) glueing} (\eg~\cite{wraith1974:artin-glueing,carboni1995:artin-glueing}), 
or \emph{change-of-base} (\eg\ \cite{jacobs:categorical-logic-and-type-theory,hermida:thesis}).
\label{exp:pullbacks-in-cat}
The objects of the pullback $\catP$ are pairs 
	$(\oC \in \catC, \oX \in \catE)$ 
such that $\funF\oC = \fp(\oX)$. We think of this as pairing $\oC$ with a generalised `relation' $\oX$. Thus, $\catP$ is the model obtained by `glueing' the models $\catC$ and $\catE$.

\begin{example}
	\label{ex:BinPred}
	We define the category of binary predicates $\cat{BPred}$ by pullback from the fibration $\fp : \Pred \to \Set$, as shown below. The objects are triples $(\oX, \oY, \oR \subset \oX \times \oY)$ and morphisms $(\mf, \mg) : (\oX, \oY, \oR) \to (\oX', \oY', \oR')$ are pairs of set-maps preserving the relation: if $(x, y) \in \oR$ then $(\mf x, \mg y) \in \oR'$. 
	\[
	%\label{eq:lifting-for-stlc}
	\begin{tikzcd}
		{\cat{BPred}} & {\cat{Pred}} \\
		\Set \times \Set & \Set
		\arrow[from=1-1, to=1-2]
		\arrow["\fq"', from=1-1, to=2-1]
		\arrow["\lrcorner"{anchor=center, pos=0.125}, draw=none, from=1-1, to=2-2]
		\arrow["\fp", from=1-2, to=2-2]
		\arrow["\times"', from=2-1, to=2-2]
	\end{tikzcd}
	\]
	Since $\fq$ is an STLC fibration then---as in \Cref{ex:logical-relations-for-stlc}---we obtain a notion of (binary) logical relations for STLC. 
	%
%	Such relations relate the interpretations in each component of the model $\Set \times \Set$. 
%	This is the first step towards \emph{effect simulation}: see \Cref{sec:effect-simulation}. 
	%
%	One can define $n$-ary relations for any natural number $n$ in the same fashion.
%	%
%	Alimohamed~\cite{alimohamed:a-characterisaton-of-lambda-definability} uses similar techniques to characterise the definable morphisms in any model of STLC. 
\end{example}

This approach extends smoothly to the effectful setting. Let us call a \emph{\monML-model} a pair $(\catC, \monT)$ consisting of a cartesian closed category $\catC$ and a strong monad $\monT$ (see~\eg~\cite{moggi-monads,mcdermott2022whatmakes}). To capture the situation of \Cref{ex:logical-relations-for-monadic-metalang}, where $\fp$ strictly preserves the interpretations of terms, our notion of fibration for logical relations for \monML\ must also preserve the monadic structure. 
Say that a \emph{\monML-fibration} from a \monML-model $(\lift\catC, \lift\monT)$ to $(\catC, \monT)$ is a fibration 
	$\fp : \lift\catC\to\catC$
which strictly preserves both cartesian closed structure and monadic structure, so that
	$\fp \circ \lift\monT = \monT \circ \fp$ 
and for all $\oX \in \lift\catC$ we have:%
	\[
		\fp(\lift\mu_{\oX}) = \mu_{\fp\oX} 					\qquad 
		\fp(\lift\eta_{\oX}) = \eta_{\fp\oX} 			\qquad
		\fp(\lift\st_{\oX}) = \st_{\fp\oX}
	\]

The fibration $\fp$ in \Cref{ex:logical-relations-for-monadic-metalang} is a \monML-fibration.
More generally, say that a \emph{\monML-model morphism} 
	$(\catB, \monS) \to (\catC, \monT)$
is a \emph{strong monad morphism}
	$(\funF, \gamma)$
such that $\funF$ preserves products.
Here $\gamma$ is a natural transformation $\funF\monS \To \monT\funF$ which is compatible with the units, multiplications, and strengths (\cf~\cite[\S 3.6]{capucci2022formaltheory}). We write $\cbvcat\fo$ for the category of \monML-models and their morphisms. The superscript emphasizes that we only require preservation of first-order structure.

\Cref{res:STLC-fibrations-closed-under-pullback} now extends to the following observation, which is essentially due to Katsumata~\cite{katsumata:semantic-formulation-of-TT-lifting,katsumata:relating-computational-effects-by-TT-lifting} (see also~\cite{kks:popl2022-full-abstraction}).

\begin{lemma}
	\label{res:pullback-of-a-CBV-fibration-is-a-CBV-fibration}
	For any \monML-fibration $\fp$ and $\cbvcat\fo$-morphism $(\funF, \gamma)$, the pullback below exists in $\cbvcat\fo$. Moreover,  $\lift\catB$ is computed as the pullback in $\Cat$ and $\fq$ is a  \monML-fibration.
	\[
		\begin{tikzcd}
			{(\lift\catB, \lift\monS)} & {(\lift\catC, \lift\monT)} \\
			{(\catB, \monS)} & {(\catC, \monT)}
			\arrow[dashed, from=1-1, to=1-2]
			\arrow[dashed, "\fq"', from=1-1, to=2-1]
			\arrow["\fp", from=1-2, to=2-2]
			\arrow["{(\funF, \gamma)}"', from=2-1, to=2-2]
		\end{tikzcd}
	\]
\end{lemma}

Katsumata's \emph{$\toptop$-lifting}~\cite{katsumata:semantic-formulation-of-TT-lifting} is a special case of this result.  For any strong monad $\monT$ on a cartesian closed category and $\monT$-algebra $(\oR, r)$ there is a canonical strong monad morphism $\sigma$ from $\monT$ to the continuation monad $(- \To \oR) \To \oR$ with result type $\oR$. Setting $(\funF, \gamma) := (\id{}, \sigma)$ above yields $\toptop$-lifting. 
 
\Cref{res:pullback-of-a-CBV-fibration-is-a-CBV-fibration} is a powerful tool for constructing new semantic models. For example, it is at the heart of the characterisation of the definable morphisms in an effectful CBV model in~\cite{kks:popl2022-full-abstraction}.  
It also provides a categorical framework for proving \emph{effect simulation} results, which show two different ways of modelling the same effect satisfy a kind of bisimilarity property: see~\eg~\cite{milner:thesis-tech-report,katsumata:relating-computational-effects-by-TT-lifting}. There it is crucial that the pullback exists even though we only require preservation of first-order structure. 

\subsection{This paper: from \texorpdfstring{\monML}{monadic metalanguage} to CBPV} 
This paper extends the theory outlined above from \monML\ to CBPV. As just outlined, this requires (1) a definition of fibration for logical relations, and (2) a theorem showing how to construct new models from old ones.
Accordingly, we provide a definition of CBPV fibrations (\Cref{sec:cbpv-fibrations}) and show how to universally construct new models in a way paralleling the two lemmas above (\Cref{res:cbpv-fibration-lifting}). To validate the theory, we show how to recover particular cases from the literature and give a version of $\toptop$-lifting for CBPV (\Cref{sec:examples}). We also show how to extend Katsumata's approach to effect simulation~\cite{katsumata:relating-computational-effects-by-TT-lifting} from \monML\ to CBPV (\Cref{sec:effect-simulation}).

The main technical obstacle is that we cannot simply define a CBPV fibration to be a functor which is both a fibration and preserves the CBPV model structure. Because CBPV is a rich language, a CBPV model consists of an adjunction enriched in presheaves over a certain category of values. Morphisms of CBPV models, therefore, make use of enriched functors. So the usual definition of fibration cannot be applied. 
Nor is it straightforward to simply write a definition by hand, because there are choices in how to define the universal property (see \Cref{rem:UMP-of-CBPV-fibration}). 
To obtain a principled definition of CBPV fibration, therefore, we must look elsewhere. 

A natural first guess would be to use enriched fibrations.  However, it is not clear this works. First, morphisms of CBPV models change the category of values, and hence the base of enrichment, so it is not clear what enriching base one should choose. Second, the theory of enriched fibrations~\cite{shulman2008framedbicats,vasilakopoulou2018enrichedfibs,moeller2020monoidalgrothendieck} is motivated by quite different concerns, namely the correspondence with the Grothendieck construction. 

Our solution is to turn to 2-category theory. 
2-categories axiomatise the structure formed by categories, functors, and natural transformations. In particular, one can define a notion of fibration internally to any 2-category. To define CBPV fibrations, we first construct an appropriate 2-category (\Cref{sec:LInd-as-a-2-category}) and then identify its internal fibrations. A CBPV fibration is then such a fibration which also preserves the CBPV model structure (\Cref{sec:defining-cbpv-fibrations}). As we outline in \Cref{sec:mathematical-context}, these steps---and indeed our main theorem---are particular instances of a general approach that applies likewise to models of STLC and \monML. More generally, we conjecture that fibrations for logical relations typically arise as internal fibrations in an appropriate 2-category of models.

A benefit of our 2-categorical approach is that we can employ the well-developed theory of internal fibrations. This also provides canonical ways to construct important examples, such as versions of the codomain and subobject fibrations.
To emphasise the value of this abstraction we also prove a relative conservativity result using an adaptation of Lafont's argument~\cite{lafont1987thesis,crole1994catsfortypes} (\Cref{sec:conserve}). This shows that, in the language without sum types, function types are a conservative extension of the first-order language. 
We view this as laying the technical groundwork for future proofs of definability and normalisation in the style of Fiore~\cite{fiore:semantic-analysis-of-nbe}.

\subsubsection*{Related work}

We believe this work represents the first full denotational account of logical relations for CBPV. Unsurprisingly, however, logical relations-style arguments have been studied for CBPV as long as the language has existed. Indeed, Levy uses a logical relations argument in~\cite[\S3.3]{levy2001thesis} (see also \cite[\S2.3]{levy2003book}), and the theory has been studied extensively from an operational perspective. 
For example, building on the work of~\cite{turi-plotkin1997:gsos} and prior work on higher-order mathematical operational semantics~\cite{goncharov2024:fossacs,goncharov2024:bialgebraic}, Goncharov, Tsampas \& Urbat~\cite{goncharov2025:cbpv} give a theory of logical relations for CBPV which includes sophisticated techniques such as step-indexing.

On the more denotational side, Kammar~\cite[Chapter 9]{kammar:thesis} gives a detailed semantic study of logical relations for CBPV, and hence effect simulation, using algebra models.   McDermott~\cite{mcdermott2020thesis} presents a denotational perspective for the more sophisticated setting of graded CBPV. His account is slightly more general than Kammar's, in that he only asks for a certain form of lifting (Definition 4.3.4, \ibid). Motivated by this account, he also shows how to define logical relations in the traditional style as relations on sets of terms (Figure 5.5, \ibid). Azevedo de Amorim~\cite{azevedodeamorim:denotational-foundations-for-expected-cost-analysis} 
presents logical relations for reasoning about the soundness of his expected cost semantics by phrasing it as an effect simulation property for a CBPV metalanguage.
%\\
These developments are particular examples of our theory. In particular, our account covers not just algebra models but the more general adjunction models as well.

\mypara{Notation.} 
%	We assume familiarity with Grothendieck fibrations: for an introduction, see~\eg~\cite{jacobs:categorical-logic-and-type-theory}. 
	We assume some basic enriched category theory, as in \eg\ \cite[Chapter 6]{borceux:handbook-of-cat-alg-vol-two}.
	We write $\catC\arr$ for the arrow category, which has objects maps in $\catC$ and morphisms commuting squares, and $\Sub\, \catC$ for the full subcategory of monomorphisms.
	In both cases we write $\cod$ for the codomain functor into $\catC$.
	If $\fp : \catE \to \catB$ is a fibration, we denote the products and exponentials in $\catE$ by $\ast$ and $\horseshoe$.
%arrow category, subobject fibration, notation for both, codomain fibration
	We assume throughout that all fibrations are split.
	Finally, because we work with enrichment in presheaf categories, size issues are relevant, especially in \Cref{sec:conserve}. Following~\cite{lucyshynwright2025:V-graded}, we handle these by assuming a hierarchy of universes of sets when required.
	\forarxivversion{}
\section{2-category theory}
\label{sec:2-categories}

We assume the basics of 2-category theory, in particular the definition of 2-categories, 2-functors, and transformations.
\forarxivversion{A brief summary is in Appendix \ref{app:2-category-theory}.}
For a detailed introduction, see 
\eg~\cite{leinster2004:book,johnson-yau2021:2-dim-cats-book}.
To fix notation, recall that a category is \emph{cartesian} if it has finite products, and a cartesian category is \emph{distributive} if it has finite coproducts and the canonical morphism 
	$[\oX \times \inj_1, \oX \times \inj_2]
		: (\oX \times \oB) + (\oX \times \oC) \to \oX \times (\oB + \oC)$
is invertible. A \emph{bicartesian closed category} (or \emph{biCCC}) is a cartesian closed category with finite coproducts. 
$\Cat$ is then the 2-category of categories. $\CartCat$ and $\DistCat$ are the 2-categories of cartesian categories and distributive categories, respectively; the 1-cells are functors preserving the structure up to isomorphism. We call such functors \emph{cartesian} and \emph{bicartesian}, respectively. The 2-cells are all natural transformations.
We write $\CartCat\strict$ and $ \DistCat\strict$ for the sub-2-categories with the same objects and functors strictly preserving the structure.

%We write $\catC, \catD, \dots$ for categories and $\C, \D, \dots$ for \mbox{2-categories}. 
The 2-categories of 2-functors $\C \to \D$ with strict (resp. pseudo / lax / oplax) natural transformations and modifications are denoted 
$
	[\C, \D]\strict,
	[\C, \D]\pseudo,
	[\C,\D]\lax$
and 
$
	[\C, \D]\oplax
$, respectively. Lax natural transformations are directed as follows:
% https://q.uiver.app/#q=WzAsNCxbMCwwLCJcXGZ1bkYgXFxvQyJdLFswLDEsIlxcZnVuR1xcb0MiXSxbMSwwLCJcXGZ1bkZcXG9DJyJdLFsxLDEsIlxcZnVuR1xcb0MnIl0sWzIsMywiXFxzaWdtYV97XFxvQyd9Il0sWzAsMiwiXFxmdW5GXFxtZiJdLFswLDEsIlxcc2lnbWFfe1xcb0N9IiwyXSxbMSwzLCJcXGZ1bkdcXG1mIiwyXSxbMiwxLCJcXHNpZ21hX1xcbWYiLDIseyJzaG9ydGVuIjp7InNvdXJjZSI6MjAsInRhcmdldCI6MzB9LCJsZXZlbCI6Mn1dXQ==
\[\begin{tikzcd}
	{\funF \oC} & {\funF\oC'} \\
	{\funG\oC} & {\funG\oC'}
	\arrow["{\funF\mf}", from=1-1, to=1-2]
	\arrow["{\sigma_{\oC}}"', from=1-1, to=2-1]
	\arrow["{\sigma_\mf}"', shorten <=5pt, shorten >=8pt, Rightarrow, from=1-2, to=2-1]
	\arrow["{\sigma_{\oC'}}", from=1-2, to=2-2]
	\arrow["{\funG\mf}"', from=2-1, to=2-2]
\end{tikzcd}\]

%We write $\C_0$ for the underlying 1-category of a 2-category~$\C$.

\subsection{Adjunctions and their morphisms}

CBPV models are defined using adjunctions internal to a 2-category. We recall the definition.

\begin{definition}
	An \emph{adjunction} in a 2-category $\C$ consists of 1-cells
		$\mf : \oA \leftrightarrows \oB : \mapu$ 
	together with 2-cells 
		$\eta : \Id{\oA} \To \mapu \circ \mf$ 
	and 
		$\epsilon : \mf \circ \mapu \To \Id{\oB}$
	satisfying the usual triangle laws. 
%	so that both composites below are the identity:
%	\[
%		\mf \xTo{\mf \circ \eta} \mf \circ \mapu \circ \mf \xTo{\epsilon \circ \mf} \mf
%		\qquad 
%		\mapu \xTo{\eta \circ \mapu} \mapu \circ \mf \circ \mapu \xTo{\epsilon \circ \mapu} \mapu
%	\]
\end{definition}

An adjunction in $\Cat$ is an adjunction in the usual sense.
%
%An adjunction in $\Cat$ is an adjunction between cartesian categories such that both functors are~cartesian. 
%
We shall also need morphisms between adjunctions. For this we shall see adjunctions as certain 2-functors and then define maps of adjunctions and their 2-cells as the corresponding transformations and modifications (\cf~\cite{pumplun1970:a-note-on-monads-and-adjoint-functors,auderset1974:adjunctions-and-monads-in-2-categories,street1986:free-adjunction}).

Let $\WAdj$ be the 2-category freely generated by the data of an adjunction, namely two objects $\bullet$ and $\ast$, 1-cells $\mf : \bullet \leftrightarrows \ast : \mapu$, and 2-cells $\eta : \Id{\bullet} \To \mapu \circ \mf$ and $\epsilon : \mf \circ \mapu \To \Id{\ast}$ satisfying the triangle laws. A 2-functor $\WAdj \to \C$ is then equivalently an adjunction in $\C$.
%
%\begin{remark}
	It follows immediately that any 2-functor preserves adjunctions.
%\end{remark}

\begin{definition}
	\label{def:adjunction-in-a-2-category}
	We write $\Adj(\C)\arb$ for the 2-functor category 
		$[\WAdj, \C]\arb$, 
	where $\arblabel \in \{ \strictlabel, \pseudolabel, \laxlabel, \oplaxlabel \}$.
	We call the 1-cells \emph{strict / pseudo / lax / oplax adjunction morphisms} and the 2-cells 
	\emph{adjunction modifications}.
\end{definition}

A lax adjunction map
	$(\ml : \oX \leftrightarrows \oY : \mr) \to (\mf : \oA \leftrightarrows \oB : \mapu)$ 
consists of 1-cells $\mm : \oX \to \oA$ and $\mn : \oY \to \oB$ with 2-cells as shown below.
The map is strict if $\alpha$ and $\beta$ are identities. 
% https://q.uiver.app/#q=WzAsOCxbMCwwLCJcXG9YIl0sWzEsMCwiXFxvWSAiXSxbMywwLCJcXG9ZICJdLFs0LDAsIlxcb1giXSxbMCwxLCJcXG9BIl0sWzEsMSwiXFxvQiJdLFszLDEsIlxcb0IiXSxbNCwxLCJcXG9BIl0sWzIsMywiXFxtciJdLFs2LDcsIlxcbWFwdSIsMl0sWzMsNywiXFxtbSJdLFsyLDYsIlxcbW4iLDJdLFswLDQsIlxcbW0iLDJdLFswLDEsIlxcbWwiXSxbMSw1LCJcXG1uIl0sWzQsNSwiXFxtZiIsMl0sWzEsNCwiXFxhbHBoYSIsMix7InNob3J0ZW4iOnsic291cmNlIjoyMCwidGFyZ2V0IjoyMH0sImxldmVsIjoyfV0sWzMsNiwiXFxiZXRhIiwyLHsic2hvcnRlbiI6eyJzb3VyY2UiOjIwLCJ0YXJnZXQiOjIwfSwibGV2ZWwiOjJ9XV0=
\[\begin{tikzcd}
	\oX & {\oY } && {\oY } & \oX \\
	\oA & \oB && \oB & \oA
	\arrow["\ml", from=1-1, to=1-2]
	\arrow["\mm"', from=1-1, to=2-1]
	\arrow["\alpha"', shorten <=4pt, shorten >=4pt, Rightarrow, from=1-2, to=2-1]
	\arrow["\mn", from=1-2, to=2-2]
	\arrow["\mr", from=1-4, to=1-5]
	\arrow["\mn"', from=1-4, to=2-4]
	\arrow["\beta"', shorten <=4pt, shorten >=4pt, Rightarrow, from=1-5, to=2-4]
	\arrow["\mm", from=1-5, to=2-5]
	\arrow["\mf"', from=2-1, to=2-2]
	\arrow["\mapu"', from=2-4, to=2-5]
\end{tikzcd}\]
The 2-cells $\alpha$ and $\beta$ must be compatible with the units and counits, in the sense that the following two diagrams commute:
% https://q.uiver.app/#q=WzAsOCxbMCwwLCJcXG1sIl0sWzEsMCwiXFxtbSBcXG1yIFxcbWwiXSxbMCwxLCJcXG1hcHUgXFxtZiBcXG1sIl0sWzEsMSwiXFxtYXB1IFxcbW4gXFxtbCJdLFszLDAsIlxcbW5cXG1sIFxcbXIgIl0sWzQsMCwiXFxtbiJdLFszLDEsIlxcbWYgXFxtbSBcXG1yIl0sWzQsMSwiXFxtZiBcXG1hcHUgXFxtbiJdLFswLDIsIlxcZXRhXntcXG1hcHUsIFxcbWZ9IFxcbWwiLDJdLFsxLDMsIlxcYmV0YSBcXG1sIl0sWzAsMSwiXFxtbCBcXGV0YV57XFxtciwgXFxtbH0iXSxbMywyLCJcXG1uIFxcYWxwaGEiXSxbNCw1LCJcXG1uXFxlcHNpbG9uXntcXG1sLCBcXG1ufSJdLFs0LDYsIlxcYWxwaGFcXG1yIiwyXSxbNiw3LCJcXG1mXFxiZXRhIiwyXSxbNyw1LCJcXGVwc2lsb25ee1xcbWYsIFxcbWFwdX0gXFxtbiIsMl1d
\begin{equation}
\label{eq:compat-law-on-lax-adjunction-maps}	
% https://q.uiver.app/#q=WzAsNCxbMCwwLCJcXG1uXFxtbCBcXG1yICJdLFsxLDAsIlxcbW4iXSxbMCwxLCJcXG1mIFxcbW0gXFxtciJdLFsxLDEsIlxcbWYgXFxtYXB1IFxcbW4iXSxbMCwxLCJcXG1uXFxlcHNpbG9uXntcXG1sLCBcXG1ufSJdLFswLDIsIlxcYWxwaGFcXG1yIiwyXSxbMiwzLCJcXG1mXFxiZXRhIiwyXSxbMywxLCJcXGVwc2lsb25ee1xcbWYsIFxcbWFwdX0gXFxtbiIsMl1d
\begin{tikzcd}
	{\mn\ml \mr } & \mn \\
	{\mf \mm \mr} & {\mf \mapu \mn}
	\arrow["{\mn\epsilon^{\ml, \mr}}", from=1-1, to=1-2]
	\arrow["{\alpha\mr}"', from=1-1, to=2-1]
	\arrow["{\mf\beta}"', from=2-1, to=2-2]
	\arrow["{\epsilon^{\mf, \mapu} \mn}"', from=2-2, to=1-2]
\end{tikzcd}
\qquad
% https://q.uiver.app/#q=WzAsNCxbMCwwLCJcXG1sIl0sWzEsMCwiXFxtbSBcXG1yIFxcbWwiXSxbMCwxLCJcXG1hcHUgXFxtZiBcXG1sIl0sWzEsMSwiXFxtYXB1IFxcbW4gXFxtbCJdLFswLDIsIlxcZXRhXntcXG1hcHUsIFxcbWZ9IFxcbWwiLDJdLFsxLDMsIlxcYmV0YSBcXG1sIl0sWzAsMSwiXFxtbCBcXGV0YV57XFxtciwgXFxtbH0iXSxbMywyLCJcXG1uIFxcYWxwaGEiXV0=
\begin{tikzcd}
	\ml & {\mm \mr \ml} \\
	{\mapu \mf \ml} & {\mapu \mn \ml}
	\arrow["{\ml \eta^{\ml, \mr}}", from=1-1, to=1-2]
	\arrow["{\eta^{\mf, \mapu} \ml}"', from=1-1, to=2-1]
	\arrow["{\beta \ml}", from=1-2, to=2-2]
	\arrow["{\mn \alpha}", from=2-2, to=2-1]
\end{tikzcd}
\end{equation}
%This is a strict adjunction map when $\alpha$ and $\beta$ are identities. 
%
%The Eilenberg--Moore and Kleisli adjunctions for a monad $T$ are then the initial and terminal objects in the category with objects the adjunctions inducing $T$ and morphisms given by strict adjunction maps (\eg~\cite[\S IV.7]{cfwm}).

\subsection{Fibrations}

We shall make extensive use of fibrations internal to a \mbox{2-category}. These have been studied in great detail (\eg~\cite{street1981:conspectus-of-variable-categories,street1974:fibrations-and-yoneda-in-a-2-cat}); for a readable introduction to the theory, see~\cite{loregian-riehl-2020:categorical-notions-of-fibration}.

%Fibrations cannot be defined as directly as adjunctions, because they use a universal property. Instead, we define them \emph{representably}: that is, using the hom-functors and the traditional definition in $\Cat$.

\begin{definition}
  Let $\C$ be a $2$-category. A \emph{fibration} in $\C$ is a 1-cell 
	$\fp : \oTotal \to \oBase$
  such that 
  \begin{enumerate}
  \item 
  	For every $\oX \in \C$, the functor 
  		$\fp \circ (-) : \C(\oX, \oTotal) \to \C(\oX, \oBase)$
  	is a fibration in $\Cat$, and
  \item 
  	For every $h : \oY \to \oX$ the following defines a morphism of fibrations, in the sense that cartesian liftings are preserved
  		(see~\eg~\cite[Definition 3.1.1]{loregian-riehl-2020:categorical-notions-of-fibration}):
  	% https://q.uiver.app/#q=WzAsNCxbMCwwLCJcXHR3b0NhdChcXG9ialgsIFxcdG90YWxPKSJdLFswLDEsIlxcdHdvQ2F0KFxcb2JqWCwgXFxiYXNlTykiXSxbMSwwLCJcXHR3b0NhdChcXG9ialhbMl0sIFxcdG90YWxPKSJdLFsxLDEsIlxcdHdvQ2F0KFxcb2JqWFsyXSwgXFxiYXNlTykiXSxbMCwyLCIoLSkgXFxjaXJjIGgiXSxbMiwzLCJcXGZpYiBcXGNpcmMgKC0pIl0sWzAsMSwiXFxmaWIgXFxjaXJjICgtKSIsMl0sWzEsMywiKC0pIFxcY2lyYyBoIiwyXV0=
  	\[\begin{tikzcd}
  		{\C(\oX, \oTotal)} & {\C(\oY, \oTotal)} \\
  		{\C(\oX, \oBase)} & {\C(\oY, \oBase)}
  		\arrow["{(-) \circ h}", from=1-1, to=1-2]
  		\arrow["{\fp \circ (-)}"', from=1-1, to=2-1]
  		\arrow["{\fp \circ (-)}", from=1-2, to=2-2]
  		\arrow["{(-) \circ h}"', from=2-1, to=2-2]
  	\end{tikzcd}\]
  \end{enumerate}
  An \emph{opfibration} is (somewhat unfortunately) defined to be a fibration in $\C\co$.
  A \emph{bifibration} is a 1-cell that is both a fibration and an opfibration.
\end{definition}

Fibrations in a 2-category inherit many of the properties of fibrations in $\Cat$. For example, it is immediate that the identity is always a fibration and that fibrations are closed under composition. An (op)fibration in $\Cat$ is exactly an (op)fibration in the usual sense (see \cite[Proposition 3.6]{gray:fibred-and-cofibred-cats}). 

The next result shows that fibrations in the 2-category of algebras for a 2-monad are exactly fibrations in the base which preserve the structure. For an introduction to the powerful theory of algebraic structure on categories via 2-monads, see~\cite{lack2009companion}. For the definition of algebras, see~\eg~\cite{power1989general,blackwell1989twodimmonad}. 

\begin{proposition}
\label{res:fibrations-in-algebras}
\begin{enumerate}
\item 
	If $\monT$ is a 2-monad on a 2-category $\C$, and $(\mf, \enr\mf)$ is a pseudomorphism of $\monT$-pseudoalgebras such that $\mf$ is a fibration in $\C$, then $(\mf, \enr\mf)$ is a fibration in $\monT\text{-}\mathrm{Alg}$.
\item 
	Right adjoint 2-functors preserve fibrations.	
\end{enumerate}
Hence, $(\mf, \enr\mf)$ is a fibration in $\monT\text{-}\mathrm{Alg}$ if and only if its underlying map is a fibration. 
\end{proposition}

This theorem covers $\CartCat, \DistCat$ and similar cases.
To characterise fibrations in our particular example, we will need some simple 2-categorical limits. 
%For an extensive discussion of both limits and fibrations, see~\cite{weber2007yoneda}. 
%First, comma objects axiomatise comma categories.

%\subsection{Products, pullbacks, and comma objects}

%Products are defined in much the same way as for categories. 
%
%\begin{definition}
%	Let $\indI$ be a small indexing set. A 2-category $\C$ has \emph{$\indI$-indexed products} if for every family $(\oA_i \mid i \in \indI)$ there is an object $\prod_i \oA_i$ and a natural isomorphism of hom-categories 
%		$\C(\oX, \prod_i \oA_i) \iso \prod_i \C(\oX, \oA_i)$
%	in $[\C, \Cat]\strict$. 
%\end{definition}

\begin{definition}[{For details, see \eg\ \cite{weber2007yoneda}}]
	Let $(\oA \xto{\mf} \oC \xleftarrow{\mg} \oB)$ be a cospan in a 2-category $\C$. The 
		\emph{comma object}
		$\mf \comma \mg$
	is the universal object with a 2-cell as shown below.
%	In $\Cat$ these are just comma categories.
	\begin{equation*}
	% https://q.uiver.app/#q=WzAsNCxbMCwwLCJcXG1mIFxcY29tbWEgXFxtZyJdLFsxLDAsIlxcb0IiXSxbMSwxLCJcXG9DIl0sWzAsMSwiXFxvQSJdLFswLDEsIlxcbXEiXSxbMSwyLCJcXG1nIl0sWzAsMywiXFxtcCIsMl0sWzMsMiwiXFxtZiIsMl0sWzYsNSwiXFxsYW1iZGEiLDIseyJzaG9ydGVuIjp7InNvdXJjZSI6MzAsInRhcmdldCI6MjB9fV1d
	\begin{tikzcd}
		{\mf \comma \mg} & \oB \\
		\oA & \oC
		\arrow["\mq", from=1-1, to=1-2]
		\arrow[""{name=0, anchor=center, inner sep=0}, "\mp"', from=1-1, to=2-1]
		\arrow[""{name=1, anchor=center, inner sep=0}, "\mg", from=1-2, to=2-2]
		\arrow["\mf"', from=2-1, to=2-2]
		\arrow["\lambda"', shorten <=16pt, shorten >=10pt, Rightarrow, from=0, to=1]
	\end{tikzcd}
	\end{equation*}
	Comma objects in $\Cat$ are just comma categories.
	The \emph{pullback} of $\mg$ along $\mf$ is defined analogously, except the square must be filled by an identity.
\end{definition}

It follows from the corresponding fact in $\Cat$ that fibrations in any 2-category are closed under pullbacks.

\begin{example}
	\label{ex:comma-in-bicartcat}
	The comma object $(\funF \comma \funG)$ in $\DistCat$ is the usual comma category.  The product
		$(\oA, \oB, \mj) \times (\oA', \oB', \mj')$ is
%		$(\funF\oA \xto{\mj} \funG\oB) \times (\funF\oA' \xto{\mj'} \funG\oB')$
	\[		
			\funF(\oA \times \oA') 
				\xto{\iso} \funF\oA \times \funF\oA' 
				\xto{\mj \times \mj'} \funG\oB \times \funG\oB' 
				\xto{\iso} \funG(\oB \times \oB') 
	\]
	and coproducts are given similarly.
%	Similar remarks hold for $\DistCat$.
\end{example}

\begin{example}
	\label{ex:arrow-objects-in-a-2-cat}
	In any 2-category with comma objects the \emph{arrow object} $\oC\arr$ on $\oC$ is defined to be the comma object 
		$(\Id{\oC} \comma \Id{\oC})$.
	In $\Cat$ this is exactly the arrow category $\catC\arr$. The definition of comma objects also gives a map  
		$\cod : \oC\arr \to \oC$;
	by \cite[Theorem 2.11]{weber2007yoneda} this is always an opfibration.
\end{example}

\begin{example}
	\label{ex:pullback-in-CartCat}
	$\CartCat$ and $\DistCat$ do not have all pullbacks. 
	Indeed, the underlying 1-categories have products but do not have all equalizers, so cannot have pullbacks; hence the 2-categories cannot have them either. For example, the two maps
		$\{ \ast \} \rightrightarrows (0 \xto\iso 1)$
	from the terminal category to the walking isomorphism do not have an equalizer in either category. 
	However, if $\fp$ is a fibration which strictly preserves products then the pullback along any map exists in $\CartCat$ (\cf~\cite[Proposition 6]{katsumata:relating-computational-effects-by-TT-lifting}).
	Similar remarks apply to $\DistCat$ when $\fp$ is also a bifibration and strictly preserves coproducts.
\end{example}

\section{Denotational models of CBPV}
\label{sec:semcbpv}

We refer to Levy's extensive works~\cite{levy2003book,levy2003adjunctionmodelsENTCS,levy2006decomposing} for the syntax and semantics of CBPV. For definiteness, we use `book CBPV', namely the basic language and complex values described in Chapters 2 \& 3 of~\cite{levy2003book}. In particular, we only ask for finite sum types and finite product types.

There are several equivalent ways to phrase the data of a CBPV model (see \cite[Chapter 11]{levy2001thesis} and \cite[\S15.1]{levy2001thesis}), so we make our choice explicit.
The basic data is a \emph{\li\ adjunction}. We refer to~\cite[\S9.3]{levy2003book} for the details on \Vli[\catC] categories, \Vli[\catC] functors and \Vli[\catC] transformations for a cartesian category $\cartesian{\catC}$, and write $\VLInd[\catC]$ for the 2-category these form. 
\forarxivversion{For a brief summary, see Appendix \ref{app:locally-indexed-structure}.}
$\VLInd[\catC]$ is equivalently the 2-category $\VCat{\Psh\catC}$ of categories enriched in the presheaf category $\cartesian{\Psh\catC}$ (see \eg~\cite[\S1.2]{kelly:basic-concepts-of-enriched-category-theory}).

We denote \Vli[\catC] categories in calligraphic font, as $\C, \D, \dots$. Maps over $\oc \in \catC$ are denoted $\oA \lito{\oc} \oB$ and the category of maps over 
	$\oc$ by $\C_\oc$. 
The reindexing functor $\C_\od \to \C_\oc$ induced by $\rho : \oc \to \od$ is denoted, in a slight departure from Levy's notation, by $(-) \reind \rho$.  

\begin{example}
	\label{ex:self}
	For a biCCC $\catC$ the \Vli[\catC] category $\self \catC$ has objects as in $\catC$ and hom-presheaves
		$(\self \catC)_{\oC}(\oA, \oB) := \catC(\oC \times \oA, \oB)$. 
\end{example}

\begin{definition}
	A \emph{\Vli[\catC] adjunction} is an adjunction in $\VLInd[\catC]$. 
	This is a pair of \Vli[\catC] functors $\funF : \C \leftrightarrows \D : \funU$ with \Vli[\catC] transformations $\eta : \id{\C} \To \funU \funF$ and $\epsilon : \id{\D} \To \funF \funU$ satisfying the usual triangle equalities as composites in $\D_1$. 
\end{definition}

A CBPV model is  a \Vli[\catC] adjunction which also models the products, sums, and function types.

\begin{definition}[{\eg~\cite[\S5]{levy2003adjunctionmodelsENTCS}}]
	Let $\catC$ be a distributive category and $\C$ be a \Vli[\catC] category. 
	\begin{enumerate}
		\item 
			$\C$ has \emph{(finite) products} if for every finite family of objects 
				$\oB_1, \dots, \oB_n$
			there exists an object $\prod_{i=1}^n \oB_i \in \C$ and arrows
				$\pi_i : \prod_{i=1}^n \oB_i \lito{1} \oB_i$
			inducing an isomorphism
				$\C_\oc(\oA, \prod_{i=1}^n \oB_i)
					\iso \prod_{i=1}^n \C_\oc(\oA, \oB_i)$.
		\item 
			$\C$ has \emph{($\catC$-indexed) powers} if for every $\oc \in \catC$ and $\oB \in \C$ there exists an object $\oc \To \oB \in \C$ and an arrow 
				$\eval : (\oc \To \oB) \lito{\oc} \oB$
			inducing an isomorphism 
				$\C_{\ob \times \oc}(\oA, \oB)
					\iso \C_{\ob}(\oA, \oc \To \oB)$.
%			natural in $\oc$ and $\oA$.
		\item 
			$\catC$'s coproducts are \emph{(finitely) distributive in $\C$} if for all 
				$\oa, \ob_1, \dots, \ob_n \in \catC$ and 
			$\oA, \oB \in \C$ the following is invertible:
			\begin{align*}	
				\C_{\oa \times \sum_{i=1}^n \ob_i}(\oA, \oB)
						&\to \smallprod_{i=1}^n \ \C_{\oa \times \ob_i}(\oA, \oB) \\
				\mf &\mapsto \big( \mf \reind\, (\id{\oa} \times \inj_i) \big)_{i=1, \dots, n}
			\end{align*}
	\end{enumerate}
\end{definition}

A CBPV model is now defined by taking the appropriate universally-defined structure for each CBPV construct.

\begin{definition}[{\eg~\cite[\S5]{levy2003adjunctionmodelsENTCS}}]
	A \emph{CBPV model} consists of a distributive category $\catC$ and a \Vli[\catC] adjunction  		
		$\funF : \self \catC \leftrightarrows \C : \funU$ 
	such that $\C$ has products and powers, and the coproducts in $\catC$ are distributive in $\C$.
\end{definition}

Value terms 
	$\Gamma \vdash^{\v} \tV : \oA$
are interpreted as maps in $\catC$, \ie\ as elements of $\catC(\sem\Gamma, \sem\oA)$. Computation terms 
	$\Gamma \vdash^{\c} \tM : \compT{\oB}$
are interpreted in
	$\catC(\sem{\Gamma}, \funU{\sem{\compT{\oB}}})$. 
Stacks 
	$\Gamma \mid \compT{\oB} \vdash^{\k} \tK : \compT{\oC}$
are interpreted in $\C_{\sem\Gamma}(\sem{\compT{\oB}},\sem{\compT{\oC}})$. 

\begin{remark}
	\label{rem:force-is-invisible}
	Because we interpret computations in 
		$\catC(\sem\Gamma, \funU\sem{\compT\oB})$
	rather than the isomorphic  
		$\C_{\sem\Gamma}(\funF1, \sem{\compT\oB})$, 
	the interpretation of $\force$\! is invisible: it is the identity.
	\forlicsversion{
	 Operationally, this reflects the fact that forcing a term does not change its behaviour.
	}
\end{remark}

For the sake of exposition, in this paper we will focus on relatively simple classes of CBPV models. We refer to~\cite[\S 4.4]{levy2003adjunctionmodelsENTCS} and~\cite{levy2003book} for the details of these and many other models.

%\begin{example}
%	The trivial adjunction 
%%		$\id{} \dashv \id{}$
%		$\id{}: \self \catC \leftrightarrows \self \catC : \id{}$
%	is a CBPV model.
%\end{example}

\begin{example}[{\cite[\S7]{levy2003adjunctionmodelsENTCS}}]
	\label{ex:syntactic-model}
	The syntax of CBPV forms a model. For any \emph{signature} $\sig$ of value base types, computation base types, and operations one may freely generate a \emph{theory} and its classifying syntactic model $\Syn_{\sig}$. 
\end{example}

\begin{example}[Algebra models]
	\label{ex:algebra-models}
	Let $\catC$ be a biCCC and $\monT$ a strong monad on $\catC$. The category $\catC^\monT$ of $\monT$-algebras becomes a \Vli[\catC] category $\EMModel{\catC}{\monT}$ with maps $(\oA, \oa) \lito{\oc} (\oB, \ob)$ the right-linear morphisms $\oc \times \oA \to \oB$. 
%	with reindexing, composition, and identities as in 
%		$\self \catC$. 
	The free--forgetful adjunction then becomes a CBPV model 
		$\funF^\monT : \self\catC \leftrightarrows \EMModel{\catC}{\monT} : \funU^\monT$.
\end{example}

\begin{example}[Storage]
	\label{ex:state-model}
	Let $\biccc\catC$ be a biCCC and $\states \in \catC$ be an object of ``states''. The adjunction 
		$(-) \times \states \dashv \states \To (-)$ 
	defines a CBPV model $\self \catC \leftrightarrows \self \catC$.
\end{example}

\section{CBPV fibrations}
\label{sec:cbpv-fibrations}

In this section we introduce CBPV fibrations. Our approach closely follows the pattern for monadic models of CBV in the style of Moggi~\cite{moggi:computational-lambda-calculus-and-monads,moggi-monads} pioneered by Katsumata~\cite{katsumata:semantic-formulation-of-TT-lifting,katsumata:a-characterisation-of-lambda-def-with-sums,katsumata:relating-computational-effects-by-TT-lifting} and others (\eg\ \cite{goubault-larrecq:logical-relations-for-monadic-types,goubault-larrecq:logical-relations-for-monadic-types-csl,kammar-mcdermott:factorisation-systems-for-logical-relations,kks:popl2022-full-abstraction}). 
%
%Katsumata~\cite{katsumata:relating-computational-effects-by-TT-lifting} defines a \emph{fibration for logical relations} to be a fibration which strictly preserves both cartesian and monadic structure. To parallel this, 

We begin by defining a 2-category of \li\ categories $\LInd$ (\Cref{sec:LInd-as-a-2-category}). 
This will play the role for CBPV that $\CartCat$ plays for STLC: it collects together the models of the basic judgements for contexts, so we can isolate models of the full language as a sub-2-category.
With this in mind, we shall define \emph{\li\ fibrations} to be fibrations internal to $\LInd$ (\Cref{sec:locally-indexed-fibrations}), and define a CBPV fibration to be a morphism of \li\ adjunctions which strictly preserves the model structure and is componentwise a \li\ fibration (\Cref{sec:defining-cbpv-fibrations}). 
Along the way we shall also see that the general theory leads directly to a variety of examples, in the form of \li\ versions of the codomain and subobject fibrations.

\subsection{The 2-category \texorpdfstring{$\LInd$}{of locally indexed categories}}
\label{sec:LInd-as-a-2-category}

We begin by defining the 2-category of \li\ categories.
The objects are pairs 
	$(\catC, \C)$
consisting of a cartesian category $\catC$ and a \Vli[\catC] category $\C$.

We want morphisms between CBPV models which may have different interpretations of values, so our 1-cells can't just be morphisms in 
	$\VLInd[\catC]$
for a fixed $\catC$.
Instead, observe that any cartesian functor
	$\funf : \catC \to \catD$ 
induces a product-preserving functor 
	$\funf\precomp := (-) \circ \funf : \Psh\catD \to \Psh\catC$ 
and hence, by \emph{change of base}  (\eg~\cite[\S 6.4]{borceux:handbook-of-cat-alg-vol-two}), a 2-functor 
	$\VLInd[\catD] \to \VLInd[\catC]$ 
we also denote by $\funf\precomp$.
Explicitly, if $\D \in \VLInd[\catD]$ then $\funf\precomp\D$ has the same objects but hom-presheaves defined by
	$(\funf\precomp\D)_{\oc} := \D_{\funf \oc}$. 
Composition and identities are as in $\D$, and reindexing along $\rho$ in $\funf\precomp\D$ is the reindexing along $\funf(\rho)$ in $\D$.

We may now define a \emph{\li\ functor}
	$(\funf, \funF) : (\catC, \C) \to (\catD, \D)$
to be a cartesian functor $\funf : \catC \to \catD$ together with a \Vli[\catC] functor 
	$\funF : \C \to \funf\precomp\D$.
 This smoothly handles reindexing: a map
	$\mk \in \C_{\oc}(\oC, \oC')$
is sent to a map
	$\funF\mk \in (\funf\precomp\D)_{\oc}(\funF\oC, \funF\oC') 
		= \D_{\funf\oc}(\funF\oC, \funF\oC')$.	

The 2-cells are defined similarly. Both change-of-base and the passage from functors $\funf$ between categories to functors $\funf\precomp$ between presheaf categories are 2-functorial~\cite{eilenberg1966closedcategories,cruttwell2008thesis} so every natural transformation $\gamma : \funf \To \fung : \catC \to \catD$ defines a strict natural transformation 
	$\gamma\precomp : \fung\precomp \To \funf\precomp : \VLInd[\catD] \to \VLInd[\catC]$.
The component $(\gamma\precomp)_\C$ at $\C \in \VLInd[\catD]$ is the identity-on-objects $\VLInd[\catC]$-functor which reindexes along $\gamma$: 
\[	
	(\gamma\precomp)_\C(\oc) :=
	(\fung\precomp\C)_\oc 
		= \C_{\fung \oc} \xto{(-) \reind \gamma_{\oc}} 
		= \C_{\funf \oc}
		= (\funf\precomp\C)_\oc
\]
We define a \emph{\li\ 2-cell}
	$(\funf, \funF) \To (\fung, \funG) : (\catC, \C) \to (\catD, \D)$
to be a natural transformation 
	$\gamma : \funf \To \fung$ 
together with a \Vli[\catC] transformation 
	$\enr\gamma : \funF \To (\alpha\precomp)_\D \circ \funG$.
Concretely, $\enr\gamma$ is a family
	$\big(  \enr\gamma_\oC : \funF\oC \lito{1} \funG\oC \big)_{\oC \in \C}$
such that for any 
	$\mk : \oC \lito{\oc} \oC'$ 
in $\C$ the following diagram commutes:
 % https://q.uiver.app/#q=WzAsNCxbMCwwLCJcXGZ1bkZcXG9DIl0sWzEsMCwiXFxmdW5GXFxvQyciXSxbMCwxLCJcXGZ1bkdcXG9DIl0sWzEsMSwiXFxmdW5HXFxvQyciXSxbMCwxLCJcXGZ1bkYoXFxtaykiXSxbMSwzLCJcXGVucntcXGdhbW1hfV97XFxvQyd9IFxccmVpbmQgXFxiYW5nX3tcXG9jfSJdLFswLDIsIlxcZW5ye1xcZ2FtbWF9X3tcXG9DfSBcXHJlaW5kIFxcYmFuZ197XFxvY30iLDJdLFsyLDMsIlxcZnVuRyhcXG1rKSBcXHJlaW5kIFxcZ2FtbWFfXFxvYyIsMl0sWzQsNywiXFxtZihcXG9jKSIsMSx7InNob3J0ZW4iOnsic291cmNlIjoyMCwidGFyZ2V0IjoyMH0sInN0eWxlIjp7ImJvZHkiOnsibmFtZSI6Im5vbmUifSwiaGVhZCI6eyJuYW1lIjoibm9uZSJ9fX1dXQ==&macro_url=https%3A%2F%2Fgist.githubusercontent.com%2Fphilipsaville%2F0345b2d81898feab11c8da414f72f776%2Fraw%2Fc5ff1a954659712dd671923c7689929e15a18f96%2Flics-25.tex
 \begin{equation}
 \label{eq:naturality-in-LInd}
 \begin{tikzcd}[column sep = 3.5em]
 	{\funF\oC} & {\funF\oC'} \\
 	{\funG\oC} & {\funG\oC'}
 	\arrow[""{name=0, anchor=center, inner sep=0}, "{\funF(\mk)}", from=1-1, to=1-2]
 	\arrow["{\enr{\gamma}_{\oC} \reind \bang_{\mf\oc}}"', from=1-1, to=2-1]
 	\arrow["{\enr{\gamma}_{\oC'} \reind \bang_{\mf\oc}}", from=1-2, to=2-2]
 	\arrow[""{name=1, anchor=center, inner sep=0}, "{\funG(\mk) \reind \gamma_\oc}"', from=2-1, to=2-2]
 	\arrow["{\mf(\oc)}"{description}, draw=none, from=0, to=1]
 \end{tikzcd}
 \end{equation}
 
 \begin{notation}
 	We henceforth adopt the notation used in \eqref{eq:naturality-in-LInd}: when writing a diagram in a \li\ category, we indicate the index by writing it in the centre of the shape.
 \end{notation}

\begin{definition}
	We write $\LInd$ for the 2-category of \li\ categories, \li\ functors, and \li\ transformations. 
\end{definition}

\begin{remark}
	\label{rem:LInd-as-a-grothendieck-construction}
	This definition is canonical: abstractly, $\LInd$ is the 2-Grothendieck construction~\cite{bakovic2010groth,buckley2014fibred2cats} of the 2-functor
		$\funK : \CartCat\coop \to \TwoCAT$
	which acts on objects by 
		$\funK(\catC) := \VLInd[\catC]$	
	and on 1-cells and 2-cells by change of base.
\end{remark}

\subsection{Locally indexed fibrations}
\label{sec:locally-indexed-fibrations}

We now characterise the fibrations internal to $\LInd$. We do this using \cite[Theorem 2.7]{weber2007yoneda}, so we need to construct comma objects. 
The following two constructions follow directly from \Cref{rem:LInd-as-a-grothendieck-construction} and the fact that, as in the 1-categorical setting, the 2-Grothendieck construction for $\funK : \C\coop \to \TwoCAT$ has those limits which exist in $\C$ and each $\funK(\oA)$ and are preserved by every $\funK(\mf)$ (\cf~\cite[\S4]{gray:fibred-and-cofibred-cats}). 

\begin{construction}
	\label{ex:commas-in-LInd}
	The comma object 
		$\big(  \funf \comma \fung, \funF \comma \funG \big)$
	of a cospan
		$(\catA, \A) \xto{(\funf, \funF)} (\catC, \C) \xleftarrow{(\fung, \funG)} (\catB, \B)$
	in $\LInd$ is defined as follows. 
	The indexing category is the comma object in $\CartCat$ (\Cref{ex:comma-in-bicartcat}). 
	The objects in 
%	the \li\ category 
	$(\funF \comma \funG)$ are triples
		$\big( \oA \in \A, \oB \in \B, \mk : \funF\oA \lito{1} \funG\oB \big)$,
	while maps
		$(\oA, \oB, \mk) \lito{\mj} (\oA', \oB', \mk')$
	over 
		$\mj : \funf\oa \to \fung\ob$
	are pairs 
		$\big( \mapu : \oA \lito{\oa} \oA', \mv : \oB \lito{\ob} \oB' \big)$
	such that following commutes:
	% https://q.uiver.app/#q=WzAsNCxbMCwwLCJcXGZ1bkZcXG9BIl0sWzEsMCwiXFxmdW5GXFxvQSciXSxbMCwxLCJcXGZ1bkdcXG9CIl0sWzEsMSwiXFxmdW5HXFxvQiciXSxbMCwxLCJcXGZ1bkYoXFxtYXB1KSJdLFswLDIsIlxcbWsgXFxyZWluZCBcXGJhbmciLDJdLFsxLDMsIlxcbWsnIFxccmVpbmRcXGJhbmciXSxbMiwzLCJcXGZ1bkcoXFxtdikgXFxyZWluZCBcXG1qIiwyXSxbNCw3LCJcXG1mKFxcb2EpIiwxLHsic2hvcnRlbiI6eyJzb3VyY2UiOjIwLCJ0YXJnZXQiOjIwfSwic3R5bGUiOnsiYm9keSI6eyJuYW1lIjoibm9uZSJ9LCJoZWFkIjp7Im5hbWUiOiJub25lIn19fV1d&macro_url=https%3A%2F%2Fgist.githubusercontent.com%2Fphilipsaville%2F0345b2d81898feab11c8da414f72f776%2Fraw%2Fc5ff1a954659712dd671923c7689929e15a18f96%2Flics-25.tex
	\begin{equation*}
	%\label{eq:map-in-comma-indexed-category}
	\begin{tikzcd}[column sep = 4em]
		{\funF\oA} & {\funF\oA'} \\
		{\funG\oB} & {\funG\oB'}
		\arrow[""{name=0, anchor=center, inner sep=0}, "{\funF(\mapu)}", from=1-1, to=1-2]
		\arrow["{\mk \reind \bang_{\mf\oa}}"', from=1-1, to=2-1]
		\arrow["{\mk' \reind\bang_{\mf\oa}}", from=1-2, to=2-2]
		\arrow[""{name=1, anchor=center, inner sep=0}, "{\funG(\mv) \reind \mj}"', from=2-1, to=2-2]
		\arrow["{\mf(\oa)}"{description}, draw=none, from=0, to=1]
	\end{tikzcd}
	\end{equation*}
	Composition, identities, and reindexing are componentwise.
	
	Similarly, the pullback of the cospan above exists in $\LInd$ when the pullback 
		$\catA \times_{\catC} \catB$
	exists in $\CartCat$, and is given by restricting $\funF \comma \funG$ to the pairs $(\oA, \oB)$ such that 
		$\funF(\oA) = \funG(\oB)$.
\end{construction}

\begin{construction}
	The product $(\catC, \C) \times (\catD, \D)$ in $\LInd$ is the 
		 	$(\catC \times \catD)$-indexed category $\C \times \D$
	with objects pairs $(\oC \in \C, \oD \in \D)$ and hom-presheaves 
			$(\C \times \D)_{(\oc, \od)} := \C_\oc \times \D_\od$.
\end{construction}

Now, working through the condition in \cite[Theorem 2.7]{weber2007yoneda}		
yields the following. To simplify notation we elide the iso
	$\fp(1) \iso 1$
given by the fact $\fp$ is cartesian.

\begin{proposition}
	\label{res:fibrations-in-LInd}
	A \li\ functor $(\fp, \fP) : (\catE, \E) \to (\catB, \B)$ is a fibration in $\LInd$ if and only if 
		$\fp$ 
	is a fibration and $\fP$ satisfies the following lifting property for any
		$\mk : \oA \lito{1} \fP\oY$:
	\[
		% https://q.uiver.app/#q=WzAsMyxbMCwwLCJcXGZQKFxcb1gpIl0sWzEsMSwiXFxvQSJdLFsyLDEsIlxcZlAoXFxvWSkiXSxbMCwxLCJcXG1qIiwyXSxbMSwyLCJcXG1rIFxccmVpbmQgXFxiYW5nIiwyXSxbMCwyLCJcXGZQKFxcbXYpIiwwLHsiY3VydmUiOi0yfV0sWzUsNCwiXFxmcChcXG9lKSIsMSx7InNob3J0ZW4iOnsic291cmNlIjoyMCwidGFyZ2V0IjoyMH0sInN0eWxlIjp7ImJvZHkiOnsibmFtZSI6Im5vbmUifSwiaGVhZCI6eyJuYW1lIjoibm9uZSJ9fX1dXQ==&macro_url=https%3A%2F%2Fgist.githubusercontent.com%2Fphilipsaville%2F0345b2d81898feab11c8da414f72f776%2Fraw%2Fc5ff1a954659712dd671923c7689929e15a18f96%2Flics-25.tex
		\hspace{2mm}
		\begin{tikzcd}[column sep = 2em]
			{\fP(\oX)} \\
			& \oA & {\fP(\oY)}
			\arrow["\mapu"', from=1-1, to=2-2]
			\arrow[""{name=0, anchor=center, inner sep=0}, "{\fP(\mv)}", curve={height=-15pt}, from=1-1, to=2-3]
			\arrow[""{name=1, anchor=center, inner sep=0}, "{\mk \reind \bang}"', from=2-2, to=2-3]
			\arrow["{\fp(\oe)}"{description,yshift=-.5mm}, draw=none, from=0, to=1]
		\end{tikzcd}
		\!\!\implies\,
		% https://q.uiver.app/#q=WzAsMyxbMCwwLCJcXG9YIl0sWzEsMSwiXFxsaWZ0XFxvQSJdLFsyLDEsIlxcb1kiXSxbMCwxLCJcXGxpZnR7XFxtan0iLDIseyJzdHlsZSI6eyJib2R5Ijp7Im5hbWUiOiJkYXNoZWQifX19XSxbMSwyLCJcXGxpZnRcXG1rIFxccmVpbmQgXFxiYW5nIiwyXSxbMCwyLCJcXG12IiwwLHsiY3VydmUiOi0yfV0sWzUsNCwiXFxvZSIsMSx7InNob3J0ZW4iOnsic291cmNlIjoyMCwidGFyZ2V0IjoyMH0sInN0eWxlIjp7ImJvZHkiOnsibmFtZSI6Im5vbmUifSwiaGVhZCI6eyJuYW1lIjoibm9uZSJ9fX1dXQ==&macro_url=https%3A%2F%2Fgist.githubusercontent.com%2Fphilipsaville%2F0345b2d81898feab11c8da414f72f776%2Fraw%2Fc5ff1a954659712dd671923c7689929e15a18f96%2Flics-25.tex
		\begin{tikzcd}[column sep = 2em]
			\oX \\
			& {\lift\oA} & \oY
			\arrow["{\lift{\mapu}}"', dashed, from=1-1, to=2-2]
			\arrow[""{name=0, anchor=center, inner sep=0}, "\mv", curve={height=-15pt}, from=1-1, to=2-3]
			\arrow[""{name=1, anchor=center, inner sep=0}, "{\lift\mk \reind \bang}"', from=2-2, to=2-3]
			\arrow["\oe"{description, yshift=-.5mm}, draw=none, from=0, to=1]
		\end{tikzcd}
	\]
	Thus, there exists $\lift\oA \in \E$ and
		$\lift\mk : \lift\oA \lito{1} \oY$
	in $\E$ such that, for any triangle in $\fp\precomp\B$ as on the left above, there exists a unique lift $\lift\mapu$ making the triangle on the right commute in $\E$.
\end{proposition}

\begin{definition}
	\label{def:locally-indexed-fibrations}
	A \emph{\li\ fibration / opfibration / bifibration} is a fibration / opfibration / bifibration in $\LInd$. 
\end{definition}

\begin{remark}
	\label{rem:UMP-of-CBPV-fibration}
	 A priori there are many possible choices for defining \li\ fibrations. If one were giving a definition by hand, one might be tempted allow maps over any index to have a cartesian lift, or specify that the unique arrow $\lift\mapu$ must be over 1.
	 Because it is derived from the mathematical theory, our definition is canonical and immediately satisfies useful properties like closure under pullback.
	 %
%	  It also fits into the schema of \li\ universal properties, in which the universal arrow lies over 1
%	  	(\cf\ \cite[Definition 11.5]{levy2003book}).
\end{remark}

We now use our mathematical framework to define \li\ versions of the core building blocks for constructing new models, namely the codomain and subobject fibrations.
The construction of the codomain opfibration follows directly from \Cref{ex:commas-in-LInd} and \Cref{ex:arrow-objects-in-a-2-cat}.

\begin{construction}
	\label{ex:pullback-in-LInd}
	The \emph{\li\ arrow category} of $(\catC, \C) \in \LInd$ is the $\catC\arr$-indexed category with objects arrows 
		$\oA \lito{1} \oB$
	in $\C_1$. There is a canonical \li\ \emph{codomain} opfibration 
		$(\cod, \cod) : (\catC\arr, \C\arr) \to (\catC, \C)$.
\end{construction}

This leads naturally to the subobject fibration. 

\begin{definition}
	We write $\Sub(\C)$  for the $\Sub(\catC)$-indexed category obtained by restricting the objects of $\C\arr$ to arrows $\oA \lito{1} \oB$ that are monic in $\C_1$. 
	Since $\Sub(\catC)$ is closed under products, reindexing is as in $\C\arr$.
\end{definition}

In $\Cat$, the codomain functor is a fibration if and only if the base category $\catC$ has pullbacks; then $\cod : \Sub\, \catC \to \catC$ is also a fibration. A corresponding fact is true here.

\begin{definition}
	A \li\ category $(\catC, \C)$ \emph{has \li\ pullbacks} 
	if $\C_1$ has pullbacks, and these are preserved by $(-) \reind \bang_\oc$ for every $\oc \in \catC$.
\end{definition}

\begin{lemma}
	Let $(\catC, \C)$ be a \li\ category. The \li\ codomain functor is a \li\ bifibration if and only if $\C$ has \li\ pullbacks. In this situation, the fibration structure restricts to make 
		$(\Sub\, \catC, \Sub\, \C) \to (\catC, \C)$
	a fibration as well.
\end{lemma}

%Asking for \li\ pullbacks is not especially strong, as the next example shows.

\begin{example}
	\label{ex:locally-indexed-pullbacks}
	Suppose $\catC$ is finitely complete. Then $\self\catC$ has locally indexed pullbacks. Moreover, for any strong monad 
		$\monT$ 
	on $\catC$ the \Vli[\catC] category $\EMModel{\catC}{\monT}$ of $\monT$-algebras also has locally indexed pullbacks.
\end{example}

\subsection{The 2-category of CBPV models}
\label{sec:CBPV-as-a-2-category}

With \li\ fibrations in hand, it remains to define a 2-category $\cbpvcat\fo\lax$  of CBPV models.
In this section we isolate $\cbpvcat\fo\lax$ as a sub-2-category of $\Adj(\LInd)\lax$. In \Cref{sec:defining-cbpv-fibrations} we will combine this with \Cref{def:locally-indexed-fibrations} to define CBPV fibrations.
We begin by defining preservation-of-structure. 

\begin{definition}
	A \Vli[\catC] functor 
		$\funF : \C \to \D$
	\emph{preserves finite products} if for any $n \in \Nat$ the canonical map
		$\tup{\funF\pi_i}_{i} : \funF(\smallprod_{i=1}^n \oC_i) \xto[1]{} \prod_{i=1}^n \funF(\oC_i)$
	is an isomorphism in $\D_1$.
	It preserves products \emph{strictly} if all the structure is preserved on the nose, so that
		$\funF(\smallprod_{i=1}^n \, \oC_i) = \smallprod_{i=1}^n\, \funF\oC_i$,
		$\funF\pi_i = \pi_i$,
	and
		$\funF(\tup{\mf_i}_i) = \tup{\funF\mf_i}_i$.
	The definition of (strict) preservation of powers is likewise.
\end{definition}

The 1-cells in $\cbpvcat\fo\lax$ are the ones we shall pull back along in our lifting theorem (\Cref{res:cbpv-fibration-lifting}), so we will only ask for preservation of first-order structure. This matches the situation for CBV, where one
only needs preservation of products to construct new models from old
	(recall \Cref{res:pullback-of-a-CBV-fibration-is-a-CBV-fibration}).

We have one more constraint to impose. A CBPV model is special kind of object in $\Adj(\LInd)\lax$: the domain of the left adjoint is of the form $\self\catC$ and the adjunction is over a single base. We therefore want 1-cells in $\cbpvcat\fo\lax$ to be maps in $\Adj(\LInd)\lax$ whose first component is determined by their action on the values.
For this we use the following lemma.

\begin{lemma}
	\label{res:self-is-a-2-functor}
	The map $\catC \mapsto \self\catC$ extends to 2-functors 
		$\CartCat \to \LInd$
	and 
		$\DistCat \to \LInd$
	which preserve products, comma objects, pullbacks, and fibrations whose underlying functor is strict cartesian (resp. cartesian and cocartesian). 
	We denote these both by $\selfname$.
\end{lemma}
%\begin{proof}[Proof notes]
%	If $\funf : \catC \to \catD$ is a bicartesian functor, then we get a \li\ functor 
%		$(\funf, \enr\funf)$
%	where
%		$\enr{\funf}(\oa) := \funf(\oa)$
%	and
%		$\enr{\funf}(\oa \xrightarrow[\oc]{\mj} \ob)$
%	is defined to be 
%	\[
%		\mf\oc \times \mf\oa \xto{\iso} \mf(\oc \times \oa) \xto{\mf(\mj)} \mf\ob
%	\]
%	The action on transformations is similar.
%\end{proof}

We now give the definition for 
	$\arblabel \in \{ \laxlabel, \oplaxlabel, \pseudolabel \}$.
The objects of $\cbpvcat\fo\arb$ are CBPV models $(\catC, \C, \funF, \funU)$.
A 1-cell $(\catC, \C, \funF^\C, \funU^\C) \to (\catD, \D, \funF^\D, \funU^\D)$ consists of
\begin{itemize}
\item 
	A bicartesian functor $\funh : \catC \to \catD$,
\item 
	A \Vli[\catC] functor 
		$\funH : \C \to \funh\precomp\D$, and
\item 
	Locally $\catC$-indexed transformations 
		$\alpha$ and $\beta$,
\end{itemize}
such that $\funH$ preserves products and
	$\big( \self\mh, \funH, (\id, \alpha), (\id, \beta) \big)$
is a 1-cell in $\Adj(\LInd)\arb$. 
A 2-cell
	$(\funh, \funH, \alpha, \beta) \To (\funh', \funH', \alpha', \beta')$ 
is a pair
	$\big( \gamma : \funh \To \funh', \enr\gamma : \funH \To \funH' \big)$ 
such that 
	$(\self \gamma, \enr\gamma)$
is a 2-cell in $\Adj(\LInd)\arb$.

In the lax case this means that 1-cells look like
% https://q.uiver.app/#q=WzAsNixbMCwwLCJcXHNlbGYgXFxjYXRDIl0sWzEsMCwiXFxDIl0sWzIsMCwiXFxzZWxmXFxjYXRDIl0sWzAsMSwiXFxzZWxmXFxjYXREIl0sWzEsMSwiXFxEIl0sWzIsMSwiXFxzZWxmXFxjYXREIl0sWzAsMSwiKFxcaWR7XFxjYXRDfSwgXFxmdW5GXlxcQykiXSxbMSwyLCIoXFxpZHtcXGNhdEN9LCBcXGZ1blVeXFxDKSJdLFswLDMsIlxcc2VsZlxcZnVuaCIsMl0sWzMsNCwiKFxcaWR7XFxjYXREfSwgXFxmdW5GXlxcRCkiLDJdLFsxLDQsIihcXGZ1bmgsIFxcZnVuSCkiLDFdLFs0LDUsIihcXGlke1xcY2F0RH0sIFxcZnVuVV5cXEQpIiwyXSxbMiw1LCJcXHNlbGZcXGZ1bmgiXSxbNywxMSwiKFxcaWQsIFxcYmV0YSkiLDAseyJvZmZzZXQiOi0xLCJzaG9ydGVuIjp7InNvdXJjZSI6MzAsInRhcmdldCI6MzB9fV0sWzYsOSwiKFxcaWQsIFxcYWxwaGEpIiwyLHsib2Zmc2V0IjotMSwic2hvcnRlbiI6eyJzb3VyY2UiOjMwLCJ0YXJnZXQiOjMwfX1dXQ==&macro_url=https%3A%2F%2Fgist.githubusercontent.com%2Fphilipsaville%2F0345b2d81898feab11c8da414f72f776%2Fraw%2Fc5ff1a954659712dd671923c7689929e15a18f96%2Flics-25.tex
\begin{equation}
\label{eq:map-of-LInd-adjunctions}
\begin{tikzcd}[column sep = 4em]
	{\self \catC} & \C & {\self\catC} \\
	{\self\catD} & \D & {\self\catD}
	\arrow[""{name=0, anchor=center, inner sep=0}, "{(\id{\catC}, \funF^\C)}", from=1-1, to=1-2]
	\arrow["{\self\funh}"', from=1-1, to=2-1]
	\arrow[""{name=1, anchor=center, inner sep=0}, "{(\id{\catC}, \funU^\C)}", from=1-2, to=1-3]
	\arrow["{(\funh, \funH)}"{description}, from=1-2, to=2-2]
	\arrow["{\self\funh}", from=1-3, to=2-3]
	\arrow[""{name=2, anchor=center, inner sep=0}, "{(\id{\catD}, \funF^\D)}"', from=2-1, to=2-2]
	\arrow[""{name=3, anchor=center, inner sep=0}, "{(\id{\catD}, \funU^\D)}"', from=2-2, to=2-3]
	\arrow["{(\id, \alpha)}"', shift left, shorten <=6pt, shorten >=6pt, Rightarrow, from=0, to=2]
	\arrow["{(\id, \beta)}", shift left, shorten <=6pt, shorten >=6pt, Rightarrow, from=1, to=3]
\end{tikzcd}
\end{equation}
so for each $\oc \in \catC$ and $\oC \in \C$ we have arrows
\[
	\alpha_\oc : (\funH\funF^\C)\oc \lito{1} (\funF^\D\funh)\oc
	\qquad
	\beta_\oC : (\funh\funU^\C)\oC \lito{1} (\funU^\D\funH)\oC
\]
natural in the sense of \eqref{eq:naturality-in-LInd} and satisfying the compatibility axioms \eqref{eq:compat-law-on-lax-adjunction-maps} as composites over 1.

%It follows essentially by construction that there is a canonical 2-functor 
%	$\cbpvcat\fo\lax \to \Adj(\LInd)\lax$
%which is injective on objects, 1-cells, and 2-cells.	

\subsection{Defining CBPV fibrations}
\label{sec:defining-cbpv-fibrations}

We can finally define CBPV fibrations as \li\ fibrations which strictly preserve structure.  Note that we require $\fp$ to be a \emph{bi}fibration because of the sum types; this is needed so that pullbacks along $\fp$ exist in $\DistCat$. Without sum types, it would be sufficient to ask for just a fibration.

\begin{definition}
	A $\cbpvcat\fo\lax$ 1-cell $(\funh, \funH, \alpha, \beta)$ is \emph{strict} if
	\begin{enumerate}
	\item 
		$\funh$ strictly preserves bicartesian structure,
	\item 
		$\funH$ strictly preserves products and powers, and
	\item 
		$(\funh, \funH)$ is a 1-cell in $\Adj(\LInd)\strict$,~\ie\ $\alpha$ and $\beta$ are both the identity.
	\end{enumerate}
	This is a \emph{CBPV (op)fibration} if $(\funh, \funH)$  is a \li\ (op)fibration and $\funh$ is a~bifibration.
\end{definition}

\begin{example}[{Recall \Cref{ex:syntactic-model}}]
	\label{ex:free-property-of-syntactic-model}
	Levy's proof of \cite[Proposition 7.3]{levy2003adjunctionmodelsENTCS} essentially shows that for any CBPV model $(\catC, \C, \funF, \funU)$ with an interpretation of the base types and operations in a signature $\sig$ there exists a strict $\cbpvcat\fo\lax$ 1-cell 
			$\Syn_\sig \to (\catC, \C, \funF, \funU)$
	extending the interpretation of $\sig$.
	Moreover, this is unique up to isomorphism.
\end{example}

By \Cref{res:self-is-a-2-functor}, a CBPV fibration is a 1-cell in
	$[\WAdj, \LInd]\strict$
which is componentwise a fibration.

Turning now to examples, one simple class of CBPV fibrations comes via monad liftings. A particular instance of the following result has been studied by Kammar~\cite[\S9.2]{kammar:thesis}, who constructs CBPV fibrations over $\Set$ using the free lifting.

\begin{lemma}
	\label{res:monad-liftings-are-cbpv-liftings}
	Let 
		$\fp : (\lift\catC, \lift\monT) \to (\catC, \monT)$
	be a \monML-fibration. Then $\fp$ extends to a fibration 
		$\widetilde{\fp} : \lift{\catC}^{\lift\monT} \to \catC^\monT$, 
	and this makes
		$(\fp, \widetilde\fp) : 
			\EMModel{\lift\catC}{\lift\monT} 
			\to 
			\EMModel{\catC}{\monT}$
	a CBPV fibration.
\end{lemma}

A further set of examples corresponds to the classical fact that, if $\catC$ is a cartesian closed category with pullbacks, then the codomain fibration over $\catC$ is an STLC fibration.
Our corresponding result is the following.

\begin{lemma}
	\label{res:lifting-structure-to-cod}
	Let $\catC$ be a cartesian category with pullbacks and $\C$ be a \Vli[\catC] category with \li\ pullbacks, products, and powers. Then $\C\arr$ and $\Sub\, \C$ both have products and powers, and the codomain \li\ functors strictly preserve this structure.
\end{lemma}

Now, the $(-)\arr$ operation is 2-functorial so from a CBPV model 
	$(\catC, \C, \funF, \funU)$ 
we obtain a lifted \Vli[\catC\arr] adjunction 
	$(\catC\arr, \C\arr, \funF\arr, \funU\arr)$. 
Combining the preceding lemma with the observation that
	$(\self \catC)\arr \iso \self (\catC\arr)$
in $\VLInd[\catC\arr]$, we obtain the following. 

\begin{proposition}
	\label{res:codomain-is-a-CBPV-model}
	Let $(\catC, \C, \funF, \funU)$ be a CBPV model such that 
		$\catC$ has pullbacks and $\C$ has \li\ pullbacks.
	Then the codomain functor 
		$(\cod, \cod) : (\catC\arr, \C\arr) \to (\catC, \C)$
	is a CBPV fibration.
\end{proposition}

The only obstacle to applying a similar argument to the subobject fibration is that the left adjoint $\funF$ may not preserve monics, and therefore may not restrict to a \li\ functor 
	$\self (\Sub\, \catC) \to \Sub\, \C$
(right adjoints always preserve monics).
We expect this can be rectified by taking an appropriate factorisation system, in the style of~\cite{goubault-larrecq:logical-relations-for-monadic-types-csl,hughes2003factorrisationsystem,goubault-larrecq:logical-relations-for-monadic-types,kammar-mcdermott:factorisation-systems-for-logical-relations}.
For reasons of space, however, we content ourselves to the case when $\funF$ preserves monics. This turns out to be common: for example, it applies to the storage model of \Cref{ex:state-model}), the erratic choice and continuation models of~\cite[\S5.7]{levy2003adjunctionmodelsENTCS}, and any algebra model over $\Set$
	(see~\cite[p. 89-90]{linton1969coeqsinalgebras}).

\begin{corollary}
	\label{res:subobject-is-a-cbpv-fibration}
	In the situation of \Cref{res:codomain-is-a-CBPV-model}, if $\funF_1$ also preserves monics then the codomain functor 
			$(\cod, \cod) : (\Sub\, \catC, \Sub\, \C) \to (\catC, \C)$
	is a CBPV fibration.
\end{corollary}

\begin{example} 
	\label{ex:fibration-on-state-model}
	Consider the model of \Cref{ex:state-model} in the case where $\catC$ has pullbacks.
	Since both the left and right adjoints preserve monics, this lifts to a storage model 
		$(-) \ast \overline\states \dashv \overline\states \horseshoe (-)$
	on 	
		$\Sub\, \catC$ 
	for any subobject 
		$\overline\states \rightarrowtail \states$.
	Then the subobject \li\ fibration is a CBPV fibration; \Cref{res:subobject-is-a-cbpv-fibration} is the case where 
		$\overline\states
			:= (\states \xto{\id{}} \states)$.
\end{example}

\section{A lifting theorem for CBPV models}
\label{sec:liftcbpv}

As we saw in \Cref{sec:constructing-fib-for-log-rel}, for applications we want a universal way to construct new fibrations for logical relations from old ones. In this section we present our central technical result, which shows this is possible for CBPV in a manner paralleling that for STLC and \monML\
	(\cf\  \Cref{res:STLC-fibrations-closed-under-pullback,res:pullback-of-a-CBV-fibration-is-a-CBV-fibration}).
	
\begin{theorem}
	\label{res:cbpv-fibration-lifting}
	Let $(\fp, \fP)$ be a CBPV fibration and $(\funh, \funH, \alpha, \beta)$ be a $\cbpvcat\fo\lax$ 1-cell. Then the pullback shown below exists in $\cbpvcat\fo\lax$ and $(\fq, \fQ)$ is a CBPV fibration.
	% https://q.uiver.app/#q=WzAsNCxbMSwwLCIoXFxsaWZ0XFxjYXRELCBcXGxpZnRcXEQsIFxcZnVuRl57XFxsaWZ0XFxEfSwgXFxmdW5VXntcXGxpZnRcXER9KSAiXSxbMSwxLCIoXFxjYXRELCBcXEQsIFxcZnVuRl5cXEQsIFxcZnVuVV5cXEQpIl0sWzAsMSwiKFxcY2F0QywgXFxDLCBcXGZ1bkZeXFxDLCBcXGZ1blVeXFxDKSJdLFswLDAsIihcXGxpZnRcXGNhdEMsIFxcbGlmdFxcQywgXFxmdW5GXntcXGxpZnRcXEN9LCBcXGZ1blVee1xcbGlmdFxcQ30pICJdLFswLDEsIihcXGZwLCBcXGZQKSJdLFsyLDEsIihcXGZ1bmgsIFxcZnVuSCwgXFxhbHBoYSwgXFxiZXRhKSIsMl0sWzMsMiwiKFxcZnEsIFxcZlEpIiwyXSxbMywwLCIoXFxsaWZ0XFxmdW5oLCBcXGxpZnRcXGZ1bkgsIFxcbGlmdFxcYWxwaGEsIFxcbGlmdFxcYmV0YSkiLDAseyJzdHlsZSI6eyJib2R5Ijp7Im5hbWUiOiJkYXNoZWQifX19XV0=&macro_url=https%3A%2F%2Fgist.githubusercontent.com%2Fphilipsaville%2F0345b2d81898feab11c8da414f72f776%2Fraw%2Fc5ff1a954659712dd671923c7689929e15a18f96%2Flics-25.tex
	\begin{equation}
		\label{eq:our-lifting-theorem}
		\begin{tikzcd}[column sep = 4em]
			{(\lift\catC, \lift\C, \funF^{\lift\C}, \funU^{\lift\C}) } & {(\lift\catD, \lift\D, \funF^{\lift\D}, \funU^{\lift\D}) } \\
			{(\catC, \C, \funF^\C, \funU^\C)} & {(\catD, \D, \funF^\D, \funU^\D)}
			\arrow["{(\lift\funh, \lift\funH, \lift\alpha, \lift\beta)}", dashed, from=1-1, to=1-2]
			\arrow[dashed, "{(\fq, \fQ)}"', from=1-1, to=2-1]
			\arrow["{(\fp, \fP)}", from=1-2, to=2-2]
			\arrow["{(\funh, \funH, \alpha, \beta)}"', from=2-1, to=2-2]
		\end{tikzcd}
	\end{equation}
\end{theorem}

This theorem is an instance of a general fact about lax transformations: see \Cref{sec:mathematical-context}. Here we sketch the concrete construction.  First, $\fq$ and $\fQ$ are defined as pullbacks in $\DistCat$ and $\LInd$ respectively;  these exist because $\fp$ is strict and a bifibration (\Cref{ex:pullback-in-CartCat} and \Cref{ex:pullback-in-LInd}). 
\begin{equation}
	\label{eq:our-main-theorem:constructing-the-objects}
	% https://q.uiver.app/#q=WzAsNCxbMCwwLCJcXGxpZnRcXGNhdEMiXSxbMCwxLCJcXGNhdEMiXSxbMSwxLCJcXGNhdEQiXSxbMSwwLCJcXGxpZnRcXGNhdEQiXSxbMSwyLCJcXGZ1bmgiLDJdLFswLDMsIlxcbGlmdFxcZnVuaCJdLFszLDIsIlxcZnAiXSxbMCwxLCJcXGZxIiwyXSxbMCwyLCIiLDIseyJzdHlsZSI6eyJuYW1lIjoiY29ybmVyIn19XV0=&macro_url=https%3A%2F%2Fgist.githubusercontent.com%2Fphilipsaville%2F0345b2d81898feab11c8da414f72f776%2Fraw%2Fc5ff1a954659712dd671923c7689929e15a18f96%2Flics-25.tex
	\begin{tikzcd}
		{\lift\catC} & {\lift\catD} \\
		\catC & \catD
		\arrow["{\lift\funh}", from=1-1, to=1-2]
		\arrow["\fq"', from=1-1, to=2-1]
		\arrow["\lrcorner"{anchor=center, pos=0.125}, draw=none, from=1-1, to=2-2]
		\arrow["\fp", from=1-2, to=2-2]
		\arrow["\funh"', from=2-1, to=2-2]
	\end{tikzcd}
	\qquad 
	\quad
	% https://q.uiver.app/#q=WzAsNCxbMCwwLCIoXFxsaWZ0XFxjYXRDLCBcXGxpZnRcXEMpIl0sWzAsMSwiKFxcY2F0QywgXFxDKSJdLFsxLDEsIihcXGNhdEQsIFxcRCkiXSxbMSwwLCIoXFxsaWZ0XFxjYXRELCBcXGxpZnRcXEQpIl0sWzEsMiwiKFxcZnVuaCwgXFxmdW5IKSIsMl0sWzAsMywiKFxcbGlmdFxcZnVuaCwgXFxsaWZ0XFxmdW5IKSJdLFszLDIsIihcXGZwLCBcXGZQKSJdLFswLDEsIihcXGZxLCBcXGZRKSIsMl0sWzAsMiwiIiwyLHsic3R5bGUiOnsibmFtZSI6ImNvcm5lciJ9fV1d&macro_url=https%3A%2F%2Fgist.githubusercontent.com%2Fphilipsaville%2F0345b2d81898feab11c8da414f72f776%2Fraw%2Fc5ff1a954659712dd671923c7689929e15a18f96%2Flics-25.tex
	\begin{tikzcd}
		{(\lift\catC, \lift\C)} & {(\lift\catD, \lift\D)} \\
		{(\catC, \C)} & {(\catD, \D)}
		\arrow["{(\lift\funh, \lift\funH)}", from=1-1, to=1-2]
		\arrow["{(\fq, \fQ)}"', from=1-1, to=2-1]
		\arrow["\lrcorner"{anchor=center, pos=0.125}, draw=none, from=1-1, to=2-2]
		\arrow["{(\fp, \fP)}", from=1-2, to=2-2]
		\arrow["{(\funh, \funH)}"', from=2-1, to=2-2]
	\end{tikzcd}
\end{equation}
An argument similar to that for cartesian closed structure (\eg~\cite[Proposition 6]{katsumata:relating-computational-effects-by-TT-lifting}) shows $(\lift\catC, \lift\C)$ has products and powers.
We define the adjunction $(\self \lift\catC \leftrightarrows \lift\C)$ and the 2-cells $\lift\alpha$ and $\lift\beta$ in \eqref{eq:our-lifting-theorem} using the universal property of the fibrations.
Observe first that the following diagram commutes because $\mathsf{self}$ is a 2-functor and $(\fp, \fP)$ is a strict adjunction morphism:
% https://q.uiver.app/#q=WzAsNixbMCwwLCJcXHNlbGYgXFxsaWZ0XFxjYXRDIl0sWzEsMCwiXFxzZWxmXFxsaWZ0XFxjYXREIl0sWzIsMCwiKFxcbGlmdFxcY2F0RCwgXFxsaWZ0XFxEKSJdLFswLDEsIlxcc2VsZlxcY2F0QyJdLFsxLDEsIlxcc2VsZlxcY2F0RCJdLFsyLDEsIihcXGNhdEQsIFxcRCkiXSxbMCwxLCJcXHNlbGYgXFxsaWZ0XFxmdW5oIl0sWzEsMiwiXFxmdW5GXntcXGxpZnRcXER9Il0sWzAsMywiXFxzZWxmXFxmcSIsMl0sWzQsNSwiXFxmdW5GXlxcRCIsMl0sWzEsNCwiXFxzZWxmIFxcZnAiLDFdLFsyLDUsIihcXGZwLCBcXGZQKSJdLFszLDQsIlxcc2VsZiBcXGZ1bmgiLDJdXQ==&macro_url=https%3A%2F%2Fgist.githubusercontent.com%2Fphilipsaville%2F0345b2d81898feab11c8da414f72f776%2Fraw%2Fc5ff1a954659712dd671923c7689929e15a18f96%2Flics-25.tex
\[\forarxivversion{\vspace{-1mm}}
\begin{tikzcd}
	{\self \lift\catC} & {\self\lift\catD} & {(\lift\catD, \lift\D)} \\
	{\self\catC} & {\self\catD} & {(\catD, \D)}
	\arrow["{\self \lift\funh}", from=1-1, to=1-2]
	\arrow["{\self\fq}"', from=1-1, to=2-1]
	\arrow["{\funF^{\lift\D}}", from=1-2, to=1-3]
	\arrow["{\self \fp}"{description}, from=1-2, to=2-2]
	\arrow["{(\fp, \fP)}", from=1-3, to=2-3]
	\arrow["{\self \funh}"', from=2-1, to=2-2]
	\arrow["{\funF^\D}"', from=2-2, to=2-3]
\end{tikzcd}\]
For any $(\oc, \lift\od) \in \self\lift\catC$ we may therefore apply the universal property of the fibration $(\fp, \fP)$ to $\alpha_{\fq(\oc, \lift\od)} = \alpha_\oc$:
For this, fix any object $(\oC, \lift\oD) \in \lift\C$ (recall \Cref{ex:pullback-in-LInd}) and apply the universal property of the fibration to the arrow $\alpha_{\fq(\oc, \lift\od)} = \alpha_\oc$:
% https://q.uiver.app/#q=WzAsNixbMCwxLCIoXFxmdW5IXFxmdW5GXlxcQylcXG9jIl0sWzEsMSwiKFxcZnVuRl5cXERcXGZ1bmgpXFxvYyA9IChcXGZQXFxmdW5GXntcXGxpZnRcXER9XFxsaWZ0XFxmdW5oKVxcb2MiXSxbMSwwLCIoXFxmdW5GXntcXGxpZnRcXER9XFxsaWZ0XFxmdW5oKVxcb2MiXSxbMCwwLCJcXGxpZnR7XFxhbHBoYV9cXG9jfShcXGZ1bkZee1xcbGlmdFxcRH1cXGxpZnRcXGZ1bmhcXG9jKSJdLFsyLDAsIlxcbGlmdFxcRCJdLFsyLDEsIlxcRCJdLFswLDEsIlxcYWxwaGFfXFxvYyJdLFszLDIsIlxcbGlmdHtcXGFscGhhfV97XFxvY30iLDAseyJzdHlsZSI6eyJib2R5Ijp7Im5hbWUiOiJkYXNoZWQifX19XSxbNCw1LCIoXFxmcCwgXFxmUCkiXSxbMCwxLCIxIiwyLHsic3R5bGUiOnsiYm9keSI6eyJuYW1lIjoibm9uZSJ9LCJoZWFkIjp7Im5hbWUiOiJub25lIn19fV0sWzMsMiwiMSIsMix7InN0eWxlIjp7ImJvZHkiOnsibmFtZSI6Im5vbmUifSwiaGVhZCI6eyJuYW1lIjoibm9uZSJ9fX1dXQ==&macro_url=https%3A%2F%2Fgist.githubusercontent.com%2Fphilipsaville%2F0345b2d81898feab11c8da414f72f776%2Fraw%2Fc5ff1a954659712dd671923c7689929e15a18f96%2Flics-25.tex
\[\begin{tikzcd}
	{\lift{\alpha_\oc}(\funF^{\lift\D}\lift\funh\oc)} & {(\funF^{\lift\D}\lift\funh)\oc} & {\lift\D} \\
	{(\funH\funF^\C)\oc} & {(\funF^\D\funh)\oc = (\fP\funF^{\lift\D}\lift\funh)\oc} & \D
	\arrow["{\lift{\alpha}_{\oc}}", dashed, from=1-1, to=1-2]
	\arrow["1"', draw=none, from=1-1, to=1-2]
	\arrow["{(\fp, \fP)}", from=1-3, to=2-3]
	\arrow["{\alpha_\oc}", from=2-1, to=2-2]
	\arrow["1"', draw=none, from=2-1, to=2-2]
\end{tikzcd}\]
This definition extends to a \li\ functor 
	$\funK : \self \lift\catC \to (\lift\catD, \D)$,
so we may use the universal property of the pullback in \eqref{eq:our-main-theorem:constructing-the-objects} to define 
	$\funF^{\lift\C}(\oc, \lift\od)$
as the unique \li\ functor filling the next diagram:
% https://q.uiver.app/#q=WzAsNixbMCwwLCJcXHNlbGYgXFxsaWZ0XFxjYXRDIl0sWzEsMSwiKFxcbGlmdFxcY2F0QywgXFxsaWZ0XFxDKSJdLFsxLDIsIihcXGNhdEMsIFxcQykiXSxbMiwyLCIoXFxjYXRELCBcXEQpIl0sWzIsMSwiKFxcbGlmdFxcY2F0RCwgXFxsaWZ0XFxEKSJdLFswLDEsIlxcc2VsZiBcXGNhdEMiXSxbMCwxLCJcXGZ1bkZee1xcbGlmdFxcQ30iLDEseyJzdHlsZSI6eyJib2R5Ijp7Im5hbWUiOiJkYXNoZWQifX19XSxbMSwyLCIoXFxmcSwgXFxmUSkiLDFdLFsyLDMsIihcXGZ1bmgsIFxcZnVuSCkiLDJdLFsxLDQsIihcXGxpZnRcXGZ1bmgsIFxcbGlmdFxcZnVuSCkiXSxbNCwzLCIoXFxmcCwgXFxmUCkiXSxbMCw0LCJcXGZ1bksiLDAseyJjdXJ2ZSI6LTJ9XSxbMCw1LCJcXHNlbGYgXFxmcSIsMl0sWzUsMiwiXFxmdW5GXlxcQyIsMl0sWzEsMywiIiwwLHsic3R5bGUiOnsibmFtZSI6ImNvcm5lciJ9fV1d&macro_url=https%3A%2F%2Fgist.githubusercontent.com%2Fphilipsaville%2F0345b2d81898feab11c8da414f72f776%2Fraw%2Fc5ff1a954659712dd671923c7689929e15a18f96%2Flics-25.tex
\[\begin{tikzcd}
	{\self \lift\catC} \\
	{\self \catC} & {(\lift\catC, \lift\C)} & {(\lift\catD, \lift\D)} \\
	& {(\catC, \C)} & {(\catD, \D)}
	\arrow["{\self \fq}"', from=1-1, to=2-1]
	\arrow["{\funF^{\lift\C}}"{description}, dashed, from=1-1, to=2-2]
	\arrow["\funK", curve={height=-12pt}, from=1-1, to=2-3]
	\arrow["{\funF^\C}"', from=2-1, to=3-2]
	\arrow["{(\lift\funh, \lift\funH)}", from=2-2, to=2-3]
	\arrow["{(\fq, \fQ)}"{description}, from=2-2, to=3-2]
	\arrow["\lrcorner"{anchor=center, pos=0.125}, draw=none, from=2-2, to=3-3]
	\arrow["{(\fp, \fP)}", from=2-3, to=3-3]
	\arrow["{(\funh, \funH)}"', from=3-2, to=3-3]
\end{tikzcd}\]
The right adjoint $\funU^{\lift\C}$ and 2-cell $\lift\beta$ are constructed similarly.

\mypara{Lifting via opfibrations.}
As a consequence of our general theory (see \Cref{sec:mathematical-context}), \Cref{res:cbpv-fibration-lifting} has a dual, as follows.

\begin{corollary}
	\label{res:cbpv-opfibration-lifting}
	Let $(\fp, \fP)$ be a CBPV opfibration and $(\funh, \funH, \alpha, \beta)$ be a $\cbpvcat\fo\oplax$ 1-cell. Then the pullback \eqref{eq:our-lifting-theorem} exists in $\cbpvcat\fo\oplax$ and $(\fq, \fQ)$ is a CBPV opfibration.
\end{corollary}

Concretely the construction is similar to that outlined above, except $\lift\alpha$ and $\lift\beta$ are defined using \emph{op}fibration structure.

\begin{remark}
\label{rem:monad-morphism-to-oplax-adjunction-map}
	\Cref{res:cbpv-opfibration-lifting} is useful because in general a monad morphism 
	$\sigma : \monS \To \monT$ (see \eg~\cite{street:formal-theory-of-monads}) induces an \emph{oplax} adjunction morphism from the Eilenberg--Moore adjunction of $\monT$ to the Eilenberg--Moore adjunction of $\monS$~\cite{pumplun1970:a-note-on-monads-and-adjoint-functors}. In such situations \Cref{res:cbpv-opfibration-lifting} applies even though \Cref{res:cbpv-fibration-lifting} does not. For a concrete example, see \Cref{ex:list-and-powerset-effect-simulation}.
\end{remark}

\subsection{Examples}
\label{sec:examples}

In this section we sketch some simple applications of our theorem. We leave a detailed exploration of the models for elsewhere: the aim is simply to show how our theorem yields a framework for building CBPV models, just as previous results do this for STLC and CBV models (\cf~\eg~\cite{mitchell-scedrov:notes-on-sconing-and-relators,hermida:thesis,kammar-mcdermott:factorisation-systems-for-logical-relations}).

\begin{example}
	We start with the storage model as in \Cref{ex:state-model}. There is a lax adjunction morphism from the storage model 
		$(-) \times \states \dashv \states \To (-)$
	on $\catC$ to the storage model 
		$(-) \times \catC(1, \states) \dashv \catC(1, \states) \To (-)$
	on $\Set$ as follows. The functors 
		$\funh$ 
	and 
		$\funH$ 
	are both given by $\catC(1, -)$. The 2-cell $\alpha$ is the isomorphism 
		$\catC(1, - \times \states) \iso \catC(1, -) \times \catC(1, \states)$,
	and 
		$\beta_{\oA} : 
			\catC(1, \states \To \oA) \to \big( \catC(1, \states) \To \catC(1, \oA) \big)$
	sends $\mt$ to $\lambda \mapu \in \catC(1, \states) \binddot \eval \circ \tup{\mt, \mapu}$.
	The model in $\Set$ is easily lifted to $\Pred$: we take any subset
		$\overline\states \subset \catC(1, \states)$
	and consider the corresponding storage model on $\Pred$ (\Cref{ex:fibration-on-state-model}).
	Applying our construction, we get a CBPV model indexed by the category 
		$\lift\catC$
	with objects pairs 
		$\big( \oC \in \catC, \oR \subset \catC(1, \oX) \big)$. 
	Since $\mathsf{self}$ commutes with pullbacks (\Cref{res:self-is-a-2-functor}), the \li\ category must also be 
		$\self \lift\catC$.
	The lifted left and right adjoints $\lift\funF$ and $\lift\funU$ are defined by 
		$\lift\funF(\oC, \oR)
			= ( \oC\times \states, \oR \times \overline\states )$
	and 
		$\lift\funF(\oC, \oR)
					= ( \states \To \oC, \overline\states \horseshoe \oR )$
\end{example}

Next we use the universal property of the syntactic model (\Cref{ex:free-property-of-syntactic-model}) to recover a definition of CBPV logical relations in the syntactic style. More precisely, from purely semantic reasoning we recover a version of the logical relations used by McDermott~\cite[p.~114]{mcdermott2020thesis}.  
%For simplicity's sake we restrict to sets-and-relations, whereas McDermott uses Kripke relations of varying arity.)

\begin{example}
	Let $\sig$ be a single-sorted signature. Then there is an associated free monad $\monT$ on $\Set$ which supports these operations, and the algebra model 
		${\funF^\monT: \self\Set \leftrightarrows \EMModel{\Set}{\monT} : \funU^\monT}$
	is a sound model of CBPV with basic operations from $\sig$.
	Explicitly, $\monT$ sends a set $\oX$ to the set of terms generated using the basic operations with variables in $\oX$ (\cf~\cite[Remark 7.2]{levy2003adjunctionmodelsENTCS}). 
	
	Now define an interpretation of base types and operations in 
		$(\Set, \Set^\monT, \funF^\monT, \funU^\monT)$
	by setting the interpretation of a value type $\oA$ to be the set of closed value terms of type $\oA$, and the interpretation of a computation type $\compT\oA$ to be the set of closed computations of type $\compT\oA$. 
	By the free property of $\Syn_\sig$, this extends to a strict map
		$\Syn_\sig \to \big( \Set, \EMModel{\Set}{\monT}, \funF^\monT, \funU^\monT \big)$.
	
	Finally, let $\lift\monT$ be a lifting of $\monT$ to $\Pred$; for definiteness, we choose the \emph{free lifting}%
		~\cite{kammar:thesis,kammar-mcdermott:factorisation-systems-for-logical-relations}.
	Now apply \Cref{res:monad-liftings-are-cbpv-liftings} and \Cref{res:cbpv-fibration-lifting} to obtain a model $(\lift\catC, \lift\C, \lift\funF, \lift\funU)$ as shown:
	% https://q.uiver.app/#q=WzAsNCxbMCwxLCJcXFN5bihcXFNpZykiXSxbMSwxLCIoXFxzZWxmXFxTZXQgXFxsZWZ0cmlnaHRhcnJvd3MgXFxTZXReXFxtb25UKSJdLFsxLDAsIihcXHNlbGYgXFxQcmVkIFxcbGVmdHJpZ2h0YXJyb3dzIFxcUHJlZF57XFxsaWZ0XFxtb25UfSkiXSxbMCwwLCIoXFxsaWZ0XFxjYXRDLCBcXGxpZnRcXEMpIl0sWzIsMSwiKFxcY29kLCBcXGNvZCkiXSxbMCwxLCJcXGV4aXN0cyAhIiwyXSxbMywyLCIiLDAseyJzdHlsZSI6eyJib2R5Ijp7Im5hbWUiOiJkYXNoZWQifX19XSxbMywwXV0=
	\[\begin{tikzcd}
		{(\lift\catC, \lift\C, \lift\funF, \lift\funU)} & {\big( \self \Pred \leftrightarrows \EMModel{\Pred}{\lift\monT} \big)} \\
		{\Syn(\sig)} & {\big( \self\Set \leftrightarrows \EMModel{\Set}{\monT} \big)}
		\arrow[dashed, from=1-1, to=1-2]
		\arrow[dashed, from=1-1, to=2-1]
		\arrow["{(\cod, \cod)}", from=1-2, to=2-2]
		\arrow["{\exists !}"', from=2-1, to=2-2]
	\end{tikzcd}\]
	Objects in $\lift\catC$ consist of a value type $\typeA$ and a set $V_\typeA$ of closed value terms of type $\typeA$. Objects in $\lift\C$ consist of a computation type $\compT\typeA$ and a set of $C_{\compT\typeA}$ of closed terms of type $\compT\typeA$, equipped with $\lift\monT$-algebra structure. The action of the adjoints $\lift\funF$ and $\lift\funU$ in the lifted model are as follows:
	\[	
		\lift\funF(\typeA, V_\typeA) = \big( \funF\typeA, \funF^{\lift\monT}V_\typeA \big)
		\quad\quad 
		\lift\funU(\compT\typeB, C_{\compT\typeA}) 
			= \big( \funU\compT\typeB, \funU^{\lift\monT}C_{\compT\typeA} \big)
	\]
	Since $\lift\monT$ is the free lifting, $\lift\funF(\typeA, V_\typeA)$ consists of the type $\funF\typeA$ and the smallest relation containing $V_\typeA$ that is closed under $\mathsf{return}$ and the operations in $\sig$. On the other hand, 
		$\lift\funU(\compT\typeB, C_{\compT\typeA})$
	consists of the type $\funU\compT\typeB$ and the set $C_{\compT\typeA}$ with its algebra structure forgotten; this reflects the fact that $\mathsf{force}$ is invisible (recall \Cref{rem:force-is-invisible}). The action on products, sums, and function types is exactly as given by McDermott.
\end{example}

%\ps{ fix this }
%\begin{remark}
%	\label{rem:equivalent-data}
%%	Levy~\cite{levy2003adjunctionmodelsENTCS} defines a CBPV model to consist of a cartesian category $\catC$, a \Vli[\catC] category $\C$, and a ``right $\C$-module'' $\O$.
%%	%
%%	This yields an especially tight connection with the syntax and operational semantics, because every term-forming operation corresponds to a natural isomorphism. However, it is somewhat unwieldy from a denotational perspective. The definition above is equivalent: $\O$ is defined using the hom-presheaves of $\C$
%%	(see \cite[\S 11.6.1]{levy2003book} and \cite[Chapter 15]{levy2001thesis}).
%	%
%	A consequence is that the isomorphism corresponding to $\mathsf{force}$ becomes the identity; thus $\mathsf{force}$ is invisible in CBPV models as defined above. This reflects the operational fact that forcing a term does not change its behaviour.
%\end{remark}

Our final example is a version of Katsumata's $\toptop$-lifting~\cite{katsumata:semantic-formulation-of-TT-lifting}, adapted for CBPV models.
Katsumata's construction relies on the fact that for any strong monad $\monT$ and any $\monT$-algebra there is a canonical strong monad morphism into the corresponding continuation monad. 
Because monad morphisms induce adjunction morphisms contravariantly (\Cref{rem:monad-morphism-to-oplax-adjunction-map}), this approach is not immediately available for adjunction models. 
Our strategy, therefore, is to first pass from our starting CBPV model to its corresponding algebra model, and then ask for a lifting of that model via \Cref{res:monad-liftings-are-cbpv-liftings}.

%\begin{remark}
%	Let us emphasise that it is \emph{not} generally the case that the output of our construction (as in \eqref{eq:our-lifting-theorem}) is an Eilenberg--Moore model, even if the starting model is. Thus, our theory does not degenerate the Eilenberg--Moore case.
%\end{remark}

In the next example we focus on $\toptop$-lifting but the construction is parametric in this choice: the argument works verbatim for any other lifting (\eg~the free lifting~\cite{kammar:thesis,kammar-mcdermott:factorisation-systems-for-logical-relations}, codensity lifting~\cite{katsumata2018codensity} or the monadic lifting of~\cite{goubault-larrecq:logical-relations-for-monadic-types-csl,goubault-larrecq:logical-relations-for-monadic-types}).

\begin{construction}[$\toptop$-lifting for CBPV]
	Let $(\catC, \C, \funF, \funU)$ be a CBPV model in which $\catC$ is also cartesian closed. Write 
	$\monT$ for the induced (strong) monad $\funU\funF$ on $\catC$. 
	By \cite[\S11.6.2]{levy2003book} there is a strict map into the algebra model 
		$\big( \catC, \EMModel{\catC}{\monT}, \funF^\monT, \funU^\monT \big)$
	for $\monT$.
	Now fix an STLC fibration 
		$\fp : \catE \to \catC$
	and an object $\oR \in \catE$ as a \emph{lifting parameter}. Finally, let 
		$\lift\monT$
	be the $\toptop$-lifting of $\monT$ with this parameter.
	By \Cref{res:monad-liftings-are-cbpv-liftings}, we obtain a CBPV fibration 
		$\big( \catE, \EMModel{\catE}{\lift\monT}, \funF^{\lift\monT}, \funU^{\lift\monT} \big)
			\to \big( \catC, \EMModel{\catC}{\monT}, \funF^{\monT}, \funU^{\monT} \big)$
	and hence, by \Cref{res:cbpv-fibration-lifting}, a lifted model as shown below:
	% https://q.uiver.app/#q=WzAsNCxbMSwwLCJcXGJpZyggXFxjYXRFLCBcXEVNTW9kZWx7XFxjYXRFfXtcXGxpZnRcXG1vblR9LCBcXGZ1bkZee1xcbGlmdFxcbW9uVH0sIFxcZnVuVV57XFxsaWZ0XFxtb25UfSBcXGJpZykiXSxbMSwxLCJcXGJpZyggXFxjYXRDLCBcXEVNTW9kZWx7XFxjYXRDfXtcXG1vblR9LCBcXGZ1bkZee1xcbW9uVH0sIFxcZnVuVV57XFxtb25UfSBcXGJpZykiXSxbMCwxLCIoXFxjYXRDLCBcXEMsIFxcZnVuRiwgXFxmdW5VKSJdLFswLDAsIihcXGxpZnRcXGNhdEMsIFxcbGlmdFxcQywgXFxsaWZ0XFxmdW5GLCBcXGxpZnRcXGZ1blUpIl0sWzIsMV0sWzAsMSwiKFxcZnAsIFxcdGlsZGVcXGZwKSJdLFszLDAsIiIsMCx7InN0eWxlIjp7ImJvZHkiOnsibmFtZSI6ImRhc2hlZCJ9fX1dLFszLDJdXQ==
	\forarxivversion{\vspace{-3mm}}
	\begin{equation}
	\label{eq:toptop-lifting-for-cbpv}
	\begin{tikzcd}
		{(\lift\catC, \lift\C, \lift\funF, \lift\funU)} & {\big( \catE, \EMModel{\catE}{\lift\monT}, \funF^{\lift\monT}, \funU^{\lift\monT} \big)} \\
		{(\catC, \C, \funF, \funU)} & {\big( \catC, \EMModel{\catC}{\monT}, \funF^{\monT}, \funU^{\monT} \big)}
		\arrow[dashed, from=1-1, to=1-2]
		\arrow[dashed, from=1-1, to=2-1]
		\arrow["{(\fp, \tilde\fp)}", from=1-2, to=2-2]
		\arrow[from=2-1, to=2-2]
	\end{tikzcd}
	\forarxivversion{\vspace{-1mm}}
	\end{equation}
	We call this the \emph{$\toptop$-lifting} of the starting model. %with lifting parameter $\oR$.
\end{construction}

%We now present a simple example of this construction.

%% \todo{ comment from a reviewer on this example, need to work out what's right and maybe explain more: "surely it would have to map a subset of {{0,1},{1}} rather than a subset of {0,1}? Alternatively, replace {{0,1},{1}} by {0,1}". }

\begin{example}
	We construct the $\toptop$-lifting of Levy's model of erratic choice~\cite[\S5.5]{levy2003book}. Thus, in our starting model the category of values is $\Set$ and the adjunction is the Kleisli resolution 
		$\funJ : \Set \leftrightarrows \Rel : \funK$
	of the powerset monad $\powerset$.
	To lift $\powerset$ to $\Pred$, we take as our lifting parameter the $\powerset$-algebra 
		$\big((\{0,1 \}, \{1\}), \mathbb{T} \big)$
	where $(\{0,1 \}, \{1\} \subseteq \{0, 1\})$ is an object of $\Pred$ and the arrow $\mathbb{T} : \powerset(\{0,1 \}) \to \{0,1 \}$ maps $p \subset \{ 0, 1\}$ to $1$ if $1\in p$ and $0$ otherwise.
	%
%	The canonical strong monad morphism 
%		$\Psh \to \Cont_{\{ \{0,1 \}, \{ 1\} \}}$
%	into the continuation monad with result type $\{ \{0,1 \}, \{ 1\} \}$ sends 
%		$p \in \Psh(\oA)$ 
%	and a predicate 
%		$\mf : \oA \to \{ 0, 1\}$
%	to $\sum_{\oa \in \oA } \mf(\oa) \, \mp(\oa)$.
	%
	Applying $\toptop$-lifting, we obtain a strong monad 
		$\lift\powerset$ 
	on $\Pred$. This acts as
		$\lift\powerset(\oA, \oR) := (\powerset\oA, \lift\powerset\oR)$ 
	where $\mp \in \lift\powerset\oR$ if and only if for all 
		$\mf : \oX \to 2$
	satisfying $\forall x \in \oR \binddot \mf(x) = 1$ we have
		$\sum_{x \in \oX} \mf(x) \, \mp(x) = 1$.
	A direct calculation then shows that 
		$\fp \in \lift\powerset\oR$
	if and only if every $x \in \fp$ is in $\oR$.
	Applying our $\toptop$-lifting construction \eqref{eq:toptop-lifting-for-cbpv}, the resulting model is the Kleisli adjunction 
		$\Pred \leftrightarrows \Pred_{\lift\powerset}$
	for $\lift\powerset$.
\end{example}

\section{Effect Simulation}
\label{sec:effectsim}

\label{sec:effect-simulation}

The \emph{effect simulation problem}~\cite{katsumata:relating-computational-effects-by-TT-lifting} is about relating different interpretations of the same computational effect. For example, one can give semantics to non-deterministic computation using either the finite powerset monad or the list monad.
The effect simulation problem asks if these semantics are ``the same'', which one could state formally as asking if the sets of possible elements denoted by the list and powerset semantics are the same.
Katsumata has studied this problem in detail for Moggi's computational $\uplambda$-calculus \cite{katsumata:relating-computational-effects-by-TT-lifting}.
%
%In this section we use \Cref{res:cbpv-fibration-lifting} to give a semantic
%account of the effect simulation problem for languages based on CBPV.
%
As we shall see, the theory we have developed thus far means we can readily extend Katsumata's approach from CBV to CBPV. 

The key idea, which has deep roots in the history of logical relations (\eg\ \cite[\S 2.2]{milner:thesis-tech-report}), is that effect
simulation is about constructing a non-standard model over the product
of the models we are trying to relate. 
Products of CBPV models are given componentwise: for models
	$\big\{ (\catC_i, \C_i, \funF_i, \funU_i) \big\}_{i=1, \dots, n}$
we get a product model 
	$\big( \smallprod_{i=1}^n \catC_i, \smallprod_{i=1}^n \C_i, \smallprod_{i=1}^n \funF_i, \smallprod_{i=1}^n \funU_i \big)$.
This is because $\LInd$ has products and the 2-functor 
	$\prod_{i=1}^n (-)$ 
lifts to a 2-functor on $\Adj(\LInd)\lax$; since $\selfname$ also preserves products, this restricts to a product on $\cbpvcat\fo\lax$. 

Semantic effect simulation now arises from \Cref{res:cbpv-fibration-lifting} as follows. We start with two CBPV models, which for brevity we denote
	$\underline{\C}_i := (\catC_i, \C_i, \funF_i, \funU_i)$
for $i = 1, 2$, a CBPV fibration $(\fp, \fP)$, and a (lax or oplax) CBPV model 1-cell as shown:
% https://q.uiver.app/#q=WzAsMyxbMCwxLCJcXHVuZGVybGluZXtcXEN9XzEgXFx0aW1lcyBcXHVuZGVybGluZXtcXEN9XzIiXSxbMSwxLCJcXHVuZGVybGluZXtcXEN9XzEgXFx0aW1lcyBcXHVuZGVybGluZXtcXEN9XzIiXSxbMSwwLCIoXFxjYXRFLCBcXEUsIFxcZnVuRl5cXEUsIFxcZnVuVV5cXEUpIl0sWzAsMSwiKFxcZnVuaCwgXFxmdW5ILCBcXGFscGhhLCBcXGJldGEpIFxcdGltZXMgXFxpZHt9IiwyXSxbMiwxLCIoXFxmcCwgXFxmUCkiXV0=
\[\begin{tikzcd}[column sep = 4em]
	& {(\catE, \E, \funF^\E, \funU^\E)} \\
	{\underline{\C}_1 \times \underline{\C}_2} & {\underline{\C}_1 \times \underline{\C}_2}
	\arrow["{(\fp, \fP)}", from=1-2, to=2-2]
	\arrow["{(\funh, \funH, \alpha, \beta) \times \id{}}"', from=2-1, to=2-2]
\end{tikzcd}\]
The effect simulation model is then constructed by applying \Cref{res:cbpv-fibration-lifting} or \Cref{res:cbpv-opfibration-lifting}.

\begin{example}
	\label{ex:list-and-powerset-effect-simulation}
  We relate the algebra models (\Cref{ex:algebra-models}) for the finite powerset monad $\powerset\fin$ and list $\List$ monad on $\Set$. Their categories of algebras are, respectively, the category $\cat{SLat}$ of sup-semilattices and $\cat{Mon}$ of
  monoids. 
  There is a canonical strong monad morphism 
  	$\gamma : \List \to \powerset\fin$ 
  sending a list to its set of elements. This gives rise to the oplax adjunction morphism below, which in turn extends to an oplax CBPV model morphism 
  	$(\id{}, \funK, \gamma, \id{}) : \EMModel{\Set}{\powerset\fin} \to \EMModel{\Set}{\List}$
  between the two algebra models.
% https://q.uiver.app/#q=WzAsNixbMCwwLCJcXFNldCJdLFswLDEsIlxcU2V0Il0sWzEsMCwiXFxjYXR7U0xhdH0iXSxbMSwxLCJcXGNhdHtNb259Il0sWzIsMCwiXFxTZXQiXSxbMiwxLCJcXFNldCJdLFswLDIsIlxcbWF0aGNhbHtQfV9mIl0sWzIsMywiIiwwLHsic3R5bGUiOnsidGFpbCI6eyJuYW1lIjoiaG9vayIsInNpZGUiOiJib3R0b20ifX19XSxbMSwzLCJbLV0iLDJdLFswLDEsImlkIiwyXSxbMiw0LCJVIl0sWzMsNSwiVSIsMl0sWzQsNSwiaWQiXSxbMSwyLCJcXGdhbW1hIiwwLHsic2hvcnRlbiI6eyJzb3VyY2UiOjIwLCJ0YXJnZXQiOjIwfSwibGV2ZWwiOjJ9XV0=
\[\begin{tikzcd}
	\Set & {\cat{SLat}} & \Set \\
	\Set & {\cat{Mon}} & \Set
	\arrow["{\mathcal{P}\fin}", from=1-1, to=1-2]
	\arrow["\id{}"', from=1-1, to=2-1]
	\arrow["U", from=1-2, to=1-3]
	\arrow["\funK", hook', from=1-2, to=2-2]
	\arrow["\id{}", from=1-3, to=2-3]
	\arrow["\gamma", shorten <=5pt, shorten >=5pt, Rightarrow, from=2-1, to=1-2]
	\arrow["{\List}"', from=2-1, to=2-2]
	\arrow["U"', from=2-2, to=2-3]
\end{tikzcd}
\]

% diagram for
% the cospan you pullback along
%
%
%\[
%	% https://q.uiver.app/#q=WzAsMyxbMCwwLCJcXEVNTW9kZWx7XFxTZXR9e1xccG93ZXJzZXRcXGZpbn0gXFx0aW1lcyBcXEVNTW9kZWx7XFxTZXR9e1xcTGlzdH0iXSxbMSwwLCJcXEVNTW9kZWx7XFxTZXR9e1xcTGlzdH1eMiJdLFsyLDAsIlxcRU1Nb2RlbHtcXGNhdHtCUHJlZH19e1xcbGlmdFxcTGlzdH0iXSxbMCwxLCIoXFxpZHt9LCBcXGZ1bkssIFxcZ2FtbWEsIFxcaWR7fSkgXFx0aW1lcyBcXGlke30iLDJdLFsyLDEsIihcXGZxLCBcXHdpZGV0aWxkZVxcZnEpIl1d&macro_url=https%3A%2F%2Fgist.githubusercontent.com%2Fphilipsaville%2F0345b2d81898feab11c8da414f72f776%2Fraw%2Fc5ff1a954659712dd671923c7689929e15a18f96%2Flics-25.tex
%	\begin{tikzcd}[column sep = 4em]
%		{\EMModel{\Set}{\powerset\fin} \times \EMModel{\Set}{\List}} & {\EMModel{\Set}{\List}^2} & {\EMModel{\cat{BPred}}{\lift\List}}
%		\arrow["{(\id{}, \funK, \gamma, \id{}) \times \id{}}"'{yshift=-1mm}, from=1-1, to=1-2]
%		\arrow["{(\fq, \widetilde\fq)}"{yshift=-1mm}, from=1-3, to=1-2]
%	\end{tikzcd}
%\]

Next consider the product model $\EMModel{\Set}{\List}^2$.
We define a CBPV fibration into this model. By \cite[Proposition 6]{katsumata:relating-computational-effects-by-TT-lifting}, the category $\cat{BPred}$ of \Cref{ex:BinPred} is a CCC with all coproducts. It follows that the free monoid monad $\lift\List$ sends $(\oX, \oY, \oR) \in \cat{BPred}$ to $\sum_{n \in \omega} (\oX, \oY, \oR)^n$; since the fibration 
	$\fq : \cat{BPred} \to \Set^2$ 
strictly preserves products and coproducts, this is a lifting of $\List \times \List$.
Hence, by \Cref{res:monad-liftings-are-cbpv-liftings}, we get a CBPV fibration
	$(\fq, \widetilde\fq) : 
			\big( \cat{BPred}, \EMModel{\cat{BPred}}{\lift\List} \big)
			\to 
			( \Set,\EMModel{\Set}{\List} )^2$.

Now we build our model for effect simulation. Applying
	\Cref{res:cbpv-opfibration-lifting}, 
we pullback $(\fq, \widetilde\fq)$ along $(\id{}, \funK, \gamma, \id{}) \times \id{}$ to obtain a CBPV model 
	$(\cat{BPred}, \BPSLatMon, \lift\funF, \lift\funU)$.
The hom-presheaves of $\BPSLatMon$ are constructed componentwise by pullback so, in particular, the category $\BPSLatMon_1$ of arrows over the terminal object arises as the pullback shown below. 
	$\cat{BPredMon}$
has objects triples $(M, N, \oR)$ such that $M$ and $N$ are monoids and $\oR$ is a submonoid of $M \times N$, and maps given by pairs of monoid morphisms that preserve the relation.
% https://q.uiver.app/#q=WzAsNCxbMCwxLCJcXGNhdHtTTGF0fVxcdGltZXNcXGNhdHtNb259Il0sWzAsMCwiXFxjYXR7QlByZWRTTGF0TW9ufSJdLFsxLDEsIlxcY2F0e01vbn1cXHRpbWVzXFxjYXR7TW9ufSJdLFsxLDAsIlxcY2F0e0JQcmVkTW9ufSJdLFsxLDBdLFszLDJdLFswLDJdLFsxLDNdLFsxLDYsIiIsMSx7ImxldmVsIjoxLCJzdHlsZSI6eyJuYW1lIjoiY29ybmVyIn19XV0=
\[\begin{tikzcd}
	{\BPSLatMon_1} & {\cat{BPredMon}} \\
	{\cat{SLat}\times\cat{Mon}} & {\cat{Mon}\times\cat{Mon}}
	\arrow[from=1-1, to=1-2]
	\arrow[from=1-1, to=2-1]
	\arrow[from=1-2, to=2-2]
	\arrow[""{name=0, anchor=center, inner sep=0}, from=2-1, to=2-2]
	\arrow["\lrcorner"{anchor=center, pos=0.125}, draw=none, from=1-1, to=0]
\end{tikzcd}\]

$\lift\funF$ acts on objects as
  $
  	\lift{F}\big( \oX, \oY, \oR) = (\powerset\fin \oX, \List\oY, \lift\oR \big),
  $
  where for a finite set $p \subset \oX$ and list $l \in \List\oY$, $(p, l) \in \lift{R}$ if
  and only if (1) for every $x \in p$ there is an element $y$ in
  $l$ such that $(x, y) \in R$, and (2) for every
  element $y$ in $l$ there's an element $x \in p$ such that 
  	$(x,y) \in R$. 
  In this new model base types $\beta$ are interpreted as the diagonal relation
  $(\sem\beta \times \sem\beta, {=})$ over an object $\sem{\beta}$ and the semantics of closed programs of type $F\beta$ are of the shape $(\gamma(l), l)$ for some list~$l$. % over $b$.
\end{example}

\section{Relative full completeness}
\label{sec:conserve}

In this section we show how our 2-categorical perspective leads relatively easily to a proof of \emph{relative full completeness}, which establishes semantically that---absent sum types---function types are a conservative extension of the first-order fragment.
Our proof follows the classic Lafont argument~\cite{lafont1987thesis}: this argument is well-known, and has been applied in many differing situations (\eg~\cite{crole1994catsfortypes,hasegawa1999glueing,fiore2002isomorphismsintypedlambda,fiore2020relativefullcompleteness}). 
Thus, our contribution here is not the proof strategy, but showing how to construct the ingredients to feed into the proof.
Indeed, as several authors have noted~\cite{hasegawa1999glueing,fiore2002isomorphismsintypedlambda}, the proof relies on:
\begin{enumerate}
\item 
	\label{it:presheaf-model-needed}
	A suitable ``presheaf'' model and a ``nerve'' construction;
\item 
	The existence of certain comma objects (``glueing'').
\end{enumerate}

In what follows we shall outline how each of these ingredients arises for \CBPV\minus, the fragment of CBPV without sum types. The rest of the argument follows the classical pattern, as in~\eg~\cite[\S4.10]{crole1994catsfortypes} so, for reasons of space, we omit it.

\begin{remark}
	We omit sum types here because of \eqref{it:presheaf-model-needed}. We want to have a map of models given by the Yoneda embedding, but the Yoneda functor does not generally preserve coproducts. There is a natural fix, namely to restrict to product-preserving presheaves (see \eg\ \cite{fu2022lambek}), but this introduces extra subtleties. Since this section is already both technical and rather compressed, we leave this for elsewhere.
	We write $\cbpvcat\fominus\lax$ for the 2-category of \CBPV\minus\ models and their morphisms, defined analogously to $\cbpvcat\fo\lax$.
\end{remark}

As well as being of interest in its own right, we view the theory sketched here as a first step towards a semantic account of Kripke relations of varying arity for full CBPV, and thereby a characterisation of definability (\cf~\cite{alimohamed:a-characterisaton-of-lambda-definability,jung-tiuryn:a-new-characterization-of-lambda-definability,kks:popl2022-full-abstraction}) and normalisation-by-evaluation in the style of~\cite{fiore:semantic-analysis-of-nbe}. This would provide a completely-denotational counterpart to \cite{abel2019:cbpv-normalisation}.

\subsection{Presheaf \li\ categories}

We construct our presheaf models for \CBPV\minus\ using the corresponding structure in $\LInd$.
The idea is to combine the enriched presheaf construction available on each 2-category $\VLInd[\catC]$ (see \eg\ \cite[\S2.2 \& \S4.4]{kelly:basic-concepts-of-enriched-category-theory}) with the presheaf construction on $\Cat$. 

First, as a \mbox{2-category} of categories enriched in a presheaf category, each $\VLInd[\catC]$ has a $\Psh\catC$-enriched presheaf construction: for every
	$\C \in \VLInd[\catC]$
there is a $\Psh\catC$-category $\Psh{\C}$ of $\Psh\catC$-functors
	$\C\op \to \Psh\catC$.
This extends to a pseudofunctor 
	$\ePsh : \VLInd[\catC] \to \VLIND[\catC]$. 
The action on 1-cells is by left Kan extension, which determines the action on 2-cells.
Applying $\ePsh$ to $\C \in \VLInd[\catC]$ yields a presheaf-like \li\ category, but over the wrong base: it is still $\catC$-indexed. We therefore apply change-of-base and define $\psmP$ as the composite 
\begin{center}
%	\label{eq:presheaves-on-LInd}
	$\VLInd[\catC] \xto{\ePsh} \VLIND[\catC] \xto{\yon} \VLIND[\Psh\catC]$
\end{center}

Using standard enriched category-theoretic techniques, together with Levy's explicit identification of $\ePsh(\C)$~\cite[p. 184]{levy2001thesis}, we arrive at the following characterisation of this composite. 

Recall from~\eg~\cite[p. 84]{levy2003adjunctionmodelsENTCS} that, for a \Vli[\catC] category $\C$, the category $\opGr{\C}$ has objects $(\oc \in \catC, \oC \in \C)$ and morphisms $(\oc, \oC) \to (\od, \oD)$ pairs of a map 
	$\rho : \od \to \oc$ 
in $\catC$ and $\mf : \oC \lito\od \oD$ in $\C$. 

\begin{definition}
	The \emph{presheaf \li\ category} $\psmP(\catC, \C) := (\Psh\catC, \Psh\C)$ is defined as follows. The objects of $\Psh\C$ are functors 
		$\funH : \opGr{\C\op} \to \Set$
	and maps $\tau : \funH \lito{\funP} \funH'$ are families of maps
%	\begin{center}	
%		\vspace{-1mm}
		$\tau_{\oc, \oC}  : \funP(\oc) \times \funH(\oc, \oC) \to \funH'(\oc, \oC)$
%		\vspace{-1mm}
%	\end{center}
	natural in each argument.  Composition, identities, and reindexing are as in 
		$\self [\opGr{\C\op}, \Set]$.
\end{definition}

A short end calculation shows that $(\Psh\catC, \Psh\C)$ is equivalently the \Vli[\Psh\catC] category obtained by reindexing $\self [\opGr\C\op, \Set]$ along the cartesian functor 
	$\pi \circ (-) : \Psh\catC \to [\opGr\C\op, \Set]$
induced by the first projection $\pi : \opGr\C \to \catC$. 
Moreover, if $\catC$ and $\catD$ are cartesian closed categories, and 
	$\funf : \catC \to \catD$
preserves products, then 
	$\funf\precomp(\self\catD) \in \VLInd[\catC]$
has products and $\catC$-powers. Hence $\Psh\C$ has products and $\Psh\catC$-powers.

\subsection{Presheaf CBPV models}

Since pseudofunctors preserve adjunctions, $\psmP$ sends a CBPV model 
	$\funF : \self \catC \leftrightarrows \C : \funU$
to a \Vli[\Psh\catC]\ adjunction 
	$\psmP(\self \catC) \leftrightarrows \psmP\C$
in which the adjoints $\funF\shriek$ and $\funU\shriek$ are computed using the left Kan extension in $\VCat{\Psh\catC}$.
To make this into a CBPV model, observe there exists an adjunction
\[	
	[\catC\op, \Set] \leftrightarrows [\opGr(\self\catC)\op, \Set] 
\]
in which the left adjoint acts by 
	$\funP \mapsto \funP(- \times {=})$
and the right adjoint acts by 
	$\funH \mapsto \funH(-, 1)$.
Using the explicit characterisation above, one sees this extends to a \Vli[\Psh\catC] adjunction
$
	\funL : 
		\self \Psh\catC \leftrightarrows \psmP(\self \catC)
	: \funR
$.
The \emph{presheaf CBPV model} is then defined to be the composite adjunction
\begin{center}
	$\funF\shriek \circ \funL :
	\self \Psh\catC
		\leftrightarrows \psmP(\self \catC)
		\leftrightarrows \psmP\C
	: \funR \circ \funU\shriek$
\end{center}

We also obtain a Yoneda map. The first component is
	$\yon : \catC \to \Psh\catC$.
For the second component we need a \Vli[\catC]-functor 
	$\Yon : \C \to \yon\precomp(\psmP\C)$.
Another end calculation shows that
	$\yon\precomp( \psmP\C )$
is isomorphic to 
	$ [\C\op, \Psh\catC]$
in $\VLIND[\catC]$, so we define $\Yon$ to be the $\Psh\catC$-enriched Yoneda embedding: 
	$\Yon(\oC) := {\C_{-}(=, \oC)}$.
This extends to a pseudonatural transformation 
%	$\mathrm{inc} \To \ePsh$
from the inclusion 
	$\VLInd[\catC] \hookrightarrow \VLIND[\catC]$
to $\ePsh$ (\cf~\cite[Lemma 3.7]{fiore2017pseudomonad}) so there exists a pseudo adjunction map in the right square below; the left square is a strict adjunction map.
%https://q.uiver.app/#q=WzAsNixbMSwwLCJcXHlvbihbXFxvcEdyIChcXHNlbGYgXFxjYXRDKVxcb3AsIFxcUHNoKFxcY2F0QyldKSAiXSxbMCwwLCJcXHNlbGYgW1xcY2F0Q1xcb3AsIFxcU2V0XSAiXSxbMiwwLCIoXFxQc2hcXGNhdEMsIFxcUHNoXFxDKSJdLFswLDEsIlxcc2VsZiBcXGNhdEMiXSxbMSwxLCJcXHNlbGYgXFxjYXRDIl0sWzIsMSwiKFxcY2F0QywgXFxDKSJdLFsxLDAsIiIsMix7Im9mZnNldCI6LTF9XSxbMCwxLCIiLDIseyJvZmZzZXQiOi0xfV0sWzAsMiwiIiwyLHsib2Zmc2V0IjotMX1dLFsyLDAsIiIsMix7Im9mZnNldCI6LTF9XSxbMywxLCJcXHNlbGYgXFx5b24iXSxbNCwwLCIoXFx5b24sIFxcWW9uKSJdLFszLDQsIiIsMSx7ImxldmVsIjoyLCJzdHlsZSI6eyJoZWFkIjp7Im5hbWUiOiJub25lIn19fV0sWzUsMiwiXFxzZWxmIFxceW9uIiwyXSxbNCw1LCIiLDEseyJvZmZzZXQiOi0xfV0sWzUsNCwiIiwxLHsib2Zmc2V0IjotMX1dLFs2LDcsIiIsMix7ImxldmVsIjoxLCJzdHlsZSI6eyJuYW1lIjoiYWRqdW5jdGlvbiJ9fV0sWzE0LDE1LCIiLDEseyJsZXZlbCI6MSwic3R5bGUiOnsibmFtZSI6ImFkanVuY3Rpb24ifX1dLFs4LDksIiIsMix7ImxldmVsIjoxLCJzdHlsZSI6eyJuYW1lIjoiYWRqdW5jdGlvbiJ9fV0sWzE0LDksIlxcaXNvIiwxLHsic2hvcnRlbiI6eyJzb3VyY2UiOjIwLCJ0YXJnZXQiOjIwfSwic3R5bGUiOnsiYm9keSI6eyJuYW1lIjoibm9uZSJ9LCJoZWFkIjp7Im5hbWUiOiJub25lIn19fV1d
\[\begin{tikzcd}[column sep = 3em]
	{\self [\catC\op, \Set] } & {\psmP(\self \catC) } & {(\Psh\catC, \Psh\C)} \\
	{\self \catC} & {\self \catC} & {(\catC, \C)}
	\arrow[""{name=0, anchor=center, inner sep=0}, shift left=1mm, from=1-1, to=1-2]
	\arrow[""{name=1, anchor=center, inner sep=0}, shift left=1mm, from=1-2, to=1-1]
	\arrow[""{name=2, anchor=center, inner sep=0}, shift left=1mm, from=1-2, to=1-3]
	\arrow[""{name=3, anchor=center, inner sep=0}, shift left=1mm, from=1-3, to=1-2]
	\arrow["{\self \yon}", from=2-1, to=1-1]
	\arrow[equals, from=2-1, to=2-2]
	\arrow["{(\yon, \Yon)}", from=2-2, to=1-2]
	\arrow[""{name=4, anchor=center, inner sep=0}, shift left=1mm, from=2-2, to=2-3]
	\arrow["{\self \yon}"', from=2-3, to=1-3]
	\arrow[""{name=5, anchor=center, inner sep=0}, shift left=1mm, from=2-3, to=2-2]
	\arrow["\dashv"{anchor=center, rotate=-90}, draw=none, from=0, to=1]
	\arrow["\dashv"{anchor=center, rotate=-90}, draw=none, from=2, to=3]
	\arrow["\iso"{description}, draw=none, from=4, to=3]
	\arrow["\dashv"{anchor=center, rotate=-90}, draw=none, from=4, to=5]
\end{tikzcd}\]

Altogether, we have shown the next proposition.

\begin{definition}
	A \li\ functor $(\funf, \funF) : (\catC, \C) \to (\catD, \D)$ is \emph{full / faithful / fully faithful} if both $\mf$ and every functor
		$\funF_\oc : \C_\oc \to \D_{\funf\oc}$
	are full / faithful / fully faithful.
	A $\cbpvcat\fominus\lax$ 1-cell $(\funf, \funF, \alpha, \beta)$ is fully faithful if $(\funf, \funF)$ is.
\end{definition}	

\begin{proposition}
	For any CBPV model $\underline{\C}$ there is a fully faithful $\cbpvcat\fominus\pseudo$ 1-cell
		$\underline{\C}
			\to \Psh{\underline\C}$
	into the presheaf CBPV model. We denote this by $\underline{\Yon}$.
\end{proposition}

Note that $\self\funf$ is fully faithful if $\funf$ is.
The final observation about presheaves we need is the following. 

\begin{proposition}
	\label{res:nerves-in-cbpvcat}
	For any $\cbpvcat\fominus\lax$ morphism 
		$\underline{\funF} : \underline{\B} \to \underline{\C}$
	there exists a $\cbpvcat\fominus\oplax$ 1-cell 
		$\underline{\nerve{\funF}}
			:
			\underline{\C}
			\to \Psh{\underline{\B}}$
	and a $\cbpvcat\fominus\oplax$ 2-cell 
		$\Gamma : \underline{\Yon} \To \underline{\nerve{\funF}} \circ \underline{\funF}$.
\end{proposition}

Indeed, for any \li\ functor
	 $(\funf, \funF) : (\catB, \B) \to (\catC, \C)$
we obtain 
	$\nerve{\funf} : \catC \to \Psh\catB$
and 
	$\nerve{\funF} : \C \to \nerve{\funf}\precomp(\Psh\B)$
by taking 
	$\nerve{\funf}{\oc} := \catC(\funf-, \oc)$
and 
	$\nerve{\funF}(\oC) := \C_{\funf-}(=, \oC)$.
Note that $\nerve{\funf}$ preserves products because $\funf$ does.
The rest of the calculation essentially follows by unwinding the standard fact---which holds equally in the enriched setting---that $\nerve{\funf}$ is the left Kan extension of $\yon$ along $\funf$.

\subsection{Completing the proof}

Let $\Syn_\sig$ be the syntactic model over a signature $\sig$ (recall \Cref{ex:syntactic-model}). Also let $\FOSyn_\sig$ be the first-order syntactic model, with function types omitted. Both these models are free (\Cref{ex:free-property-of-syntactic-model}) so there is a canonical strict map of first-order \CBPV\minus\ models
	$(i, I) : \FOSyn_\sig \to \Syn_{\sig}$.
We prove the following. 

\begin{theorem}[Relative full completeness]
  \label{th:conservativity}
	For any signature $\sig$, the \li\ functor $(i, I)$ is fully faithful.
\end{theorem}	
	
The remaining difficulty lies in seeing that for any $\cbpvcat\fominus\oplax$ morphism
		$(\fung, \funG, \alpha, \beta)$ 
the following comma object exists, \ie\ CBPV models admit \emph{glueing}:
% https://q.uiver.app/#q=WzAsNCxbMCwxLCJcXHVuZGVybGluZXtcXEJ9Il0sWzEsMSwiXFx1bmRlcmxpbmVcXEMiXSxbMSwwLCJcXHVuZGVybGluZVxcQyJdLFswLDAsIlxcdW5kZXJsaW5le1xcR30iXSxbMywyXSxbMiwxLCIiLDAseyJsZXZlbCI6Miwic3R5bGUiOnsiaGVhZCI6eyJuYW1lIjoibm9uZSJ9fX1dLFswLDEsIihcXGZ1bmcsIFxcZnVuRywgXFxhbHBoYSwgXFxiZXRhKSIsMl0sWzMsMF0sWzUsNywiXFxsYW1iZGEiLDAseyJzaG9ydGVuIjp7InNvdXJjZSI6MzAsInRhcmdldCI6MzB9fV1d
\[\begin{tikzcd}[column sep = 5em]
	{\underline{\G}} & {\underline\C} \\
	{\underline{\B}} & {\underline\C}
	\arrow[from=1-1, to=1-2]
	\arrow[""{name=0, anchor=center, inner sep=0}, from=1-1, to=2-1]
	\arrow[""{name=1, anchor=center, inner sep=0}, equals, from=1-2, to=2-2]
	\arrow["{(\fung, \funG, \alpha, \beta)}"', from=2-1, to=2-2]
	\arrow["\lambda", shorten <=17pt, shorten >=17pt, Rightarrow, from=1, to=0]
\end{tikzcd}\]
This follows from two facts. First, for any 2-category $\diag$, if $\C$ has comma objects then 
$[\diag, \C]\oplax$ also has such comma objects, computed component-wise (\cf~\cite[Proposition 4.6]{lack2005limitsforlax}). Since $\LInd$ has all comma objects, 
so does
	$\Adj(\LInd)\oplax$. 
Second, a small adaptation of the classical proof (\eg~\cite{crole1994catsfortypes}) shows this restricts to $\cbpvcat\oplax\ppres$: when $\underline\C$ is a CBPV model with \li\ pullbacks,   $\underline\G$ is also a CBPV model.

The rest of the argument is as in the classical case (see~\eg~\cite[\S4.10]{crole1994catsfortypes} or \cite[\S3.2]{fiore2002isomorphismsintypedlambda}), observing that composition in $\cbpvcat\fominus\oplax$ reduces to (1) composition of the functors in $\DistCat$ on the first component, and (2) on the second component, composition in $\Cat$ at each index. 

\section{Lifting theorems for arbitrary shapes}
\label{sec:mathematical-context}

In this final technical section we outline how \Cref{res:cbpv-fibration-lifting} and \Cref{res:cbpv-opfibration-lifting} are special cases of a general result which applies to any ``shape'' of model, including those for \monML\ and CBPV. 
The key idea is that, since $\cbpvcat\fo\lax$ is a sub-2-category of the functor 2-category 
	$\Adj(\LInd)\lax = [\WAdj, \LInd]\lax$,
\Cref{res:cbpv-fibration-lifting} is a special case of a result about pullbacks in 2-categories of the form $[\diag, \C]\lax$ for some 2-category $\diag$ of ``diagram shapes''.
%
%Thus, \Cref{res:cbpv-fibration-lifting} is a statement about the existence of pullbacks in a 2-category of 2-functors and lax transformations. 
%
We state the general version in \Cref{res:general-lifting-theorem}.
%
%To handle the strict and lax maps at play, we implicitly follow the treatment of fibrations in~\cite{arkor2024:enhanced}. 

For the theorem we need to isolate a class of 1-cells which will play the role of fibrations for logical relations, \ie\ fibrations which strictly preserve structure. We do this in a style inspired by~\cite{arkor2024:enhanced}. Fix a 2-category $\C$ and a wide, locally-full sub-2-category $\C\tight$; we call the 1-cells in $\C\tight$ \emph{tight}. Further assume every fibration is tight.
The statement is then as follows.

\begin{theorem}
	\label{res:general-lifting-theorem}
	In the situation just sketched, consider a cospan 
		$(\funG \xto\phi \funH \xleftarrow\kappa \lift\funH)$
	in $[\diag, \C]\lax$ such that 
	\begin{enumerate}
	\item 
		$\kappa$ is strict and each 1-cell component $\kappa_\od$ is tight;
	\item 
		For each $\od \in \diag$ the pullback $(\phi_\od)\precomp(\kappa_\od)$ of $\kappa_\od$ along $\phi_\od$ exists in $\C$ and is tight.
	\end{enumerate}
	Then the 1-cells $(\phi_\od)\precomp(\kappa_\od)$ form a strict, componentwise-tight transformation, which is the pullback $\phi\precomp(\kappa)$ in $[\diag, \C]\lax$.
\end{theorem}

Essentially, this says that to construct pullbacks of models with additional structure encoded by a 2-functor it suffices to construct pullbacks of the underlying `base' model. The proof is similar to that sketched in \Cref{sec:liftcbpv}. 

We recover \Cref{res:cbpv-fibration-lifting} as follows.
A closely-related result has also been observed by Nishizawa, Katsumata \& Komorida as part of their study of Stone dualities: see \cite[Theorem 7.6]{nishizawa2020:stone}.

% for the existence of pullbacks in the co: 
% via Nathanael: C should have J-limits iff C^co has J^co-limits, immediately from the universal property. But the diagram for 2-pullbacks is invariant under (-)^co.
\begin{example}
	 Take $\diag := \WAdj$ and $\C$ to be the sub-2-category of $\LInd$ with objects 
		$(\catC, \C)$
	given by a distributive category $\catC$ and a locally $\catC$-indexed category $\C$ with finite products and $\catC$-powers, such that $\catC$'s coproducts are distributive. The 1-cells are $\LInd$-maps which preserve the products and coproducts. Say a 1-cell $(\fp, \fP)$ is tight if $\fp$ is a bifibration---so pullbacks of $\fp$ exist in $\DistCat$---and $(\fp, \fP)$ is a \li\ fibration which strictly preserves the products, coproducts, and powers. $\C$ satisfies the conditions of \Cref{res:general-lifting-theorem}, with the required pullbacks computed as in $\LInd$. Since $\selfname$ also preserves pullbacks (\Cref{res:self-is-a-2-functor}), the pullback of a CBPV model is still a CBPV model. So we obtain \Cref{res:cbpv-fibration-lifting}.
	\Cref{res:cbpv-opfibration-lifting} follows by instantiating the theorem with 
		$\LInd\co$.
\end{example}

We obtain a version for \monML\ by varying the ``shapes''.
By \cite[\S3.6]{capucci2022formaltheory}, applying the 2-Grothendieck construction to
the 2-functor
		$\CartCat\op \to \TwoCAT$ 
sending $\catV$ to the 2-category 
		$\VAct[\catV]\lax$
of $\catV$-actions and lax maps yields a 2-category $\Act$ of actions. 
The objects are triples of a cartesian category $\catV$, a category $\catC$, and a left action of $\catV$ on $\catC$.
1-cells 
	$(\catV, \catC, {\bullet}) \to  (\catW, \catD, {\star})$
are triples $(\funf, \funF, \phi)$ where $\funf: \catV\to \catW$ is cartesian, $\funF: \catC \to \catD$, and 
	$\phi_{\oV, \oC} : \funf(\oV) \star \funF(\oC) \to \funF(\oV \bullet \oC)$
is a natural transformation compatible with the two actions.
$\Act$ plays the role for \monML\ that $\LInd$ did for CBPV above. 
First, a monad internal to $\Act$ (see~\cite{street:formal-theory-of-monads}) is exactly a left action together with a monad that is strong for the action (\eg~\cite[\S3]{mcdermott2022whatmakes}). 
Second, using \cite[Theorem 2.7]{weber2007yoneda}, one sees the fibrations in $\Act$ are 1-cells 
	$(\funf, \funF, \phi)$ 
such that both $\funf$ and $\funF$ are fibrations.
Finally, there is a 2-functor $\self : \CartCat \to \Act$ sending a cartesian category $\catV$ to the canonical $\catV$-action on itself. This picks out the underlying structure of a \monML\ model, and preserves both fibrations and pullbacks (\cf\ \Cref{res:self-is-a-2-functor}).

\begin{example}
	Take $\diag := \WMnd$ to be the walking monad, \ie\ the 2-category freely generated by the data of a monad (\eg\ \cite{auderset1974:adjunctions-and-monads-in-2-categories,street1986:free-adjunction}), and let $\C$ be the sub-2-category of $\Act$ with objects $(\catV, \catC, \bullet)$ such that $\catV$ is cartesian closed.  Say a 1-cell is tight if it is an $\Act$-fibration which strictly preserves both the cartesian closed structure and the action. 
	Then consider a cospan
		$(\self \catC \xto{\self \funF} \self \catB \xleftarrow{\self \fp} \self \catE)$
	where $\fp$ is a fibration which strictly preserves the cartesian closed structure.
	Since $\fp$ is tight and the pullback exists in $\Act$, by \Cref{res:general-lifting-theorem} this becomes a pullback of \monML\ models, which yields \Cref{res:pullback-of-a-CBV-fibration-is-a-CBV-fibration}.
\end{example}

%\section{Perspectives}
%
%In this paper we have shown how fibrations internal to 2-categories
%can be used to define a mathematically robust and principled
%definintion of logical relations for CBPV. We have illustrated the
%viability of our definition by showing how familiar fibrational
%logical relations can be generalized to CBPV. As our main application,
%we use our framework to prove, for the first time, a conservativity
%property of CBPV.

%We see this example as a first step towards other metatheoretical
%properties such as definability and normalization-by-evaluation.
%Another promising direction for further work is using our framework
%to propose a fibrational account to logical relations for graded
%type theories.

\section*{Acknowledgments}

We are grateful to the Programming Languages Group in Oxford for many constructive conversations, and to Nathanael Arkor for important technical discussions.
We also thank the LICS 2025 reviewers for their useful suggestions, and Shin-ya Katsumata for pointing out \cite{nishizawa2020:stone}.

\bibliography{refs}

%\appendix
\appendices
%\onecolumn

\section{The basic definitions of 2-category theory}
\label{app:2-category-theory}

We briefly review the definitions of 2-categories, 2-functors, transformations, and modifications.
For reasons of space we omit the coherence axioms: for full details see \eg~\cite{leinster2004:book,johnson-yau2021:2-dim-cats-book}. 

\begin{definition}
	A \emph{2-category} $\C$ consists of:
	\begin{enumerate}
	\item 
		A collection of objects $\oA, \oB, \dots$.
	\item 
		For all objects $\oA$ and $\oB$, a collection of morphisms from $\oA$ to $\oB$, which are themselves related by morphisms: thus we have a \emph{hom-category} $\C(\oA, \oB)$ whose objects $\mf, \mg : \oA \to \oB$  are \emph{morphisms} (or \emph{1-cells}) and whose morphisms are \emph{2-cells} $\sigma, \tau : \mf \To \mg$. 
	\item 
		For all $\oA, \oB$ and $\oC$ a \emph{composition} functor 
			$\circ_{\oA, \oB, \oC} : \C(\oB, \C) \times \C(\oA, \oB) \to \C(\oA, \oC)$
		and, for all $\oA$ an \emph{identity} 1-cell $\Id\oA : \oA \to \oA$,
		such that composition is associative and unital on both 1-cells and 2-cells.
	\end{enumerate}
\end{definition}

The hom-category structure means the following. For any 1-cell $\mf : \oA \to \oB$ there is an identity 2-cell $\id\mf : \mf \To \mf$. Moreover, given $\sigma : \mf \To \mg : \oA \to \oB$ and $\tau : \mg \To \mh : \oA \to \oB$ we can \emph{vertically compose} to obtain a 2-cell $\tau \vert \sigma : \mf \To \mh$. The functoriality of composition says that, given $\sigma : \mf \To \mf' : \oA \to \B$ and $\tau : \mg \To \mg' : \oB \to \oC$ we can \emph{horizontally compose} to obtain a 2-cell 
	$\tau \circ \sigma : \mg \circ \mf \To \mg' \circ \mf'$,
and that the two composition operations are related by the so-called interchange law.
The names correspond to how the operations look when drawn in $\Cat$ (see \cite[\S II.5]{cfwm}).
Following standard 2-categorical practice, we sometimes write simply
	$\oA \sigma$
or 
	$\mf\sigma$
instead of 
	$\Id\oA \circ \sigma$
or 
	$\id\mf \circ \sigma$, 
and similarly for composition on the other side.

Every 2-category $\C$ has three duals, corresponding to reversing just the 1-cells, just the 2-cells, or both. $\C\op$ has $\C\op(\oA, \oB) := \C(\oB, \oA)$, so just the 1-cells are reversed. $\C\co$ has 
	$\C\co(\oA, \oB) := \C(\oA, \oB)\op$, 
so just the 2-cells are reversed. $\C\coop$ has 
		$\C\co(\oA, \oB) := \C(\oB, \oA)\op$,
so both 1-cells and 2-cells are reversed.

As indicated above, the prototypical example is $\Cat$: the objects are functors, the 1-cells are functors, and the 2-cells are natural transformations.

\begin{definition}
	A \emph{2-functor} $\funF : \C \to \D$ is a mapping on objects, 1-cells, and 2-cells which preserves both horizontal and vertical composition, so that 
		$\funF(\mg \circ \mf) = \funF(\mg) \circ \funF(\mf)$
	and 
		$\funF(\Id{\oA}) = \Id{\funF\oA}$ on 1-cells, 
	and 
		$\funF(\tau \circ \sigma) = \funF(\tau) \circ \funF(\sigma)$,
		$\funF(\tau \vert \sigma) = \funF(\tau) \vert \funF(\sigma)$,
	and 
		$\funF(\id\mf) = \id{\funF\mf}$
	on 2-cells.
\end{definition}

When considering functors between monoidal categories, there are four grades of strictness we can ask for: strict, pseudo (strong), lax, or oplax. The same applies for morphisms between 2-functors.

\begin{definition}
	Let $\funF, \funG : \C \to \D$ be 2-functors. A \emph{lax natural transformation} 
		$\sigma : \funF \to \funG$
	consists of a 1-cell $\sigma_\oC : \funF\oC \to \funG\oC$ for each $\oC\in \C$ together with, for every 1-cell $\mf : \oC \to \oC'$ in $\C$, a 2-cell 
		$\sigma_f$
	witnessing the naturality as shown below:
	% https://q.uiver.app/#q=WzAsNCxbMCwwLCJcXGZ1bkYgXFxvQyJdLFswLDEsIlxcZnVuR1xcb0MiXSxbMSwwLCJcXGZ1bkZcXG9DJyJdLFsxLDEsIlxcZnVuR1xcb0MnIl0sWzIsMywiXFxzaWdtYV97XFxvQyd9Il0sWzAsMiwiXFxmdW5GXFxtZiJdLFswLDEsIlxcc2lnbWFfe1xcb0N9IiwyXSxbMSwzLCJcXGZ1bkdcXG1mIiwyXSxbMiwxLCJcXHNpZ21hX1xcbWYiLDIseyJzaG9ydGVuIjp7InNvdXJjZSI6MjAsInRhcmdldCI6MzB9LCJsZXZlbCI6Mn1dXQ==
	\[\begin{tikzcd}
		{\funF \oC} & {\funF\oC'} \\
		{\funG\oC} & {\funG\oC'}
		\arrow["{\funF\mf}", from=1-1, to=1-2]
		\arrow["{\sigma_{\oC}}"', from=1-1, to=2-1]
		\arrow["{\sigma_\mf}"', shorten <=5pt, shorten >=8pt, Rightarrow, from=1-2, to=2-1]
		\arrow["{\sigma_{\oC'}}", from=1-2, to=2-2]
		\arrow["{\funG\mf}"', from=2-1, to=2-2]
	\end{tikzcd}\]
	This is required to satisfy unit and associativity laws, and be natural in $\mf$. An \emph{oplax} natural transformation is a transformation in $\C\co$: the 2-cell $\sigma_\mf$ is reversed. A \emph{pseudonatural} transformation is a transformation in which every $\sigma_\mf$ is an isomorphism. A \emph{strict} (or \emph{2-natural}) transformation is one in which every $\sigma_\mf$ is the identity.
\end{definition}

Transformations play the same role as natural transformations in category theory. Since 2-category theory has an extra layer of data, there is an extra form of morphism. We state the definition for lax natural transformations; corresponding definitions hold for oplax, pseudo, and strict transformations.

\begin{definition}
	Let $\sigma, \tau : \funF \to \funG : \C \to \D$ be lax natural transformations. A \emph{modification} 
		$\Gamma : \sigma \to \tau$ 
	consists of a 2-cell $\Gamma_\oC : \sigma_\oC \To \tau_\oC$ for each $\oC \in \C$, subject to an axiom making it compatible with the 2-cells $\sigma_\mf$ and $\tau_\mf$.
\end{definition}

For any 2-categories $\C$ and $\D$, we therefore obtain four 2-functor categories. For each  
	$w \in \{ \strictlabel, \pseudolabel, \laxlabel, \oplaxlabel \}$ 
there is a 2-category 
	$[\C, \D]_w$
with objects the 2-functors $\C \to \D$, 1-cells either the strict, pseudo, lax, or oplax transformations, and 2-cells the modifications.

\section{Locally indexed categories, functors, and transformations}
\label{app:locally-indexed-structure}

Throughout this section, fix $\cartesian\catC$ to be a cartesian category. We recall the following from \cite[\S 9.3]{levy2003book}.

\begin{definition}
	\label{def:locally-indexed-category}
	A \emph{\Vli[\catC] category} $\C$ consists of:
	\begin{enumerate}
	\item		
		A collection $\Ob\catC$ of objects $\oA, \oB, \dots$.
	\item 
		For each $\oc \in \catC$ a category $\C_\oc$ with objects $\Ob\catC$. Thus for each 
			$\oA, \oB \in \C$
		we have a set 
			$\C_\oc(\oA, \oB)$
		of \emph{morphisms over $\oc$}, denoted 
			$\mf : \oA \lito\oc \oB$;
	\item 
		For each $\rho : \od \to \oc$ in $\catC$ an identity-on-objects functor
			$(-) \reind \rho : \C_\oc \to \C_\od$, 
		subject to the following axioms for 
			$\rho' : \oe \to \od$,
			$\mf : \oA \lito\oc \oB$
		and 
			$\mg : \oB \lito\oc \oC$:
		\begin{align*}
			\mf \reind \id{\oc} = \mf
			\quad 
			&
			\quad
			\mf \reind (\rho \circ \rho')
				=  (\mf \reind \rho) \reind \rho'
%			\\
%			(\mg \circ \mf) \reind \rho 
%				&= (\mg \reind \rho) \circ (\mf \reind \rho)
		\end{align*}
	\end{enumerate}
\end{definition}

\begin{definition}
	A \emph{\Vli[\catC] functor} $\funF : \C \to \D$ consists of a mapping
		$\Ob\funF : \Ob\C \to \Ob\D$
	on objects and, for every $\oc \in \catC$ and $\oA, \oB \in \C$, a mapping 
		$\funF_\oc : \C_\oc(\oA, \oB) \to \D_\oc(\funF\oA, \funF\oB)$
	such that 
		$\funF_\oc$ defines a functor $\C_\oc \to \D_\oc$ and is
	compatible with renaming:
		$\funF_\oc(\mf \reind \rho) = \funF_\oc(\mf) \reind \rho$
	for any $\rho : \od \to \oc$ in $\catC$.		
	We drop the subscripts where they are clear from context.
\end{definition}

\begin{definition}
	A \emph{\Vli[\catC] transformation} $\alpha : \funF \To \funG : \C \to \D$ consists of an arrow 
		$\alpha_\oC : \funF\oC \lito{1} \funG\oC$ 
	for each $\oC \in \C$, natural in the sense that for any 
		$\mf : \oB \lito{\oc} \oC$
	in $\C$ the following diagram commutes in $\D_\oc$:
	% https://q.uiver.app/#q=WzAsNCxbMSwwLCJcXGZ1bkZcXG9DIl0sWzEsMSwiXFxmdW5HXFxvQyJdLFswLDAsIlxcZnVuRlxcb0IiXSxbMCwxLCJcXGZ1bkdcXG9CIl0sWzIsMCwiXFxmdW5GXFxtZiJdLFswLDEsIlxcYWxwaGFfQyBcXHJlaW5kIFxcYmFuZ19cXG9jIl0sWzIsMywiXFxhbHBoYV9CIFxccmVpbmQgXFxiYW5nX1xcb2MiLDJdLFszLDEsIlxcZnVuR1xcbWYiLDJdLFs0LDcsIjEiLDEseyJzaG9ydGVuIjp7InNvdXJjZSI6MjAsInRhcmdldCI6MjB9LCJzdHlsZSI6eyJib2R5Ijp7Im5hbWUiOiJub25lIn0sImhlYWQiOnsibmFtZSI6Im5vbmUifX19XV0=&macro_url=https%3A%2F%2Fgist.githubusercontent.com%2Fphilipsaville%2F0345b2d81898feab11c8da414f72f776%2Fraw%2Fc5ff1a954659712dd671923c7689929e15a18f96%2Flics-25.tex
	\begin{equation}
	\label{eq:locally-ind-trans}
	\begin{tikzcd}
		{\funF\oB} & {\funF\oC} \\
		{\funG\oB} & {\funG\oC}
		\arrow[""{name=0, anchor=center, inner sep=0}, "{\funF\mf}", from=1-1, to=1-2]
		\arrow["{\alpha_B \reind \bang_\oc}"', from=1-1, to=2-1]
		\arrow["{\alpha_C \reind \bang_\oc}", from=1-2, to=2-2]
		\arrow[""{name=1, anchor=center, inner sep=0}, "{\funG\mf}"', from=2-1, to=2-2]
		\arrow["\oc"{description}, draw=none, from=0, to=1]
	\end{tikzcd}
	\end{equation}
\end{definition}

\begin{remark}
	Maps over 1 play an important role in \Vli[\catC] categories. This is because they are categories enriched in $\Psh\catC$: the monoidal unit in $\Psh\catC$ is 
		$\yon(1) := \catC(-, 1)$ 
	so, by the Yoneda lemma, morphisms from 
		$\yon(1)$
	into the hom-presheaf $\C_{(-)} : \catC\op \to \Set$ are in bijection with maps over~1.
\end{remark}

\end{document}

%% file: macros.tex
\newcommand{\To}{\Rightarrow}
\newcommand{\id}[1]{\mathrm{id}_{#1}}
\newcommand{\inj}{\mathrm{inj}}
\newcommand{\eval}{\mathrm{eval}}
\newcommand{\yon}{y}
\newcommand{\Yon}{Y}
\newcommand{\nerve}[1]{\tup{#1}}
\newcommand{\op}{^{\mathrm{op}}}
\newcommand{\co}{^{\mathrm{co}}}
\newcommand{\coop}{^{\mathrm{coop}}}
\newcommand{\iso}{\cong}

\newcommand{\Ob}[1]{| #1 |}
\newcommand{\comma}{\,{\downarrow}\,}

\newcommand{\Nat}{\mathbb{N}}

\newcommand{\Psh}[1]{\widehat{#1}}
\newcommand{\ePsh}{\mathrm{P}}
\newcommand{\Adj}{\mathrm{Adj}}

\newcommand{\WMnd}{\mathbf{Mnd}}
\newcommand{\WAdj}{\mathbf{Adj}}
\newcommand{\powerset}{\mathcal{P}}
\newcommand{\psmP}{\mathcal{P}} % presheaf pseudomonad

\newcommand{\Set}{\cat{Set}}
\newcommand{\Rel}{\cat{Rel}}
\newcommand{\Cat}{\cat{Cat}}
\newcommand{\CAT}{\cat{CAT}}

\newcommand{\TwoCAT}{2\text{-}\CAT}
\newcommand{\VCat}[1]{{#1}\text{-}\cat{Cat}}
\newcommand{\VLInd}[1][\catV]{{#1}\text{-}\mathbf{LInd}}
\newcommand{\VLIND}[1][\catV]{{#1}\text{-}\mathbf{LIND}}
\newcommand{\LInd}{\mathbf{LInd}}

\newcommand{\CartCat}{\cat{CartCat}}

\newcommand{\DistCat}{\cat{DistCat}}

\newcommand{\Act}{\cat{Act}}

\newcommand{\VAct}[1][\catV]{{#1}\text{-}\cat{Act}}
\newcommand{\Pred}{\cat{Pred}}

\newcommand{\BPSLatMon}{\mathcal{SLML}\mathit{ift}} %{\mathcal{BPSL}\mathit{at}\mathcal{M}\mathit{on}}

\newcommand{\minus}{$^{-}$}
\newcommand{\ppres}{^{\!{\times}}}

\newcommand{\fo}{^{{\mathrm{fo}}}}
\newcommand{\fominus}{^{{\mathrm{-,fo}}}}
\newcommand{\tight}{_{\mathit{t}}}

\newcommand{\arblabel}{\mathit{w}}

\newcommand{\strictlabel}{\mathrm{st}}
\newcommand{\pseudolabel}{\mathrm{ps}}
\newcommand{\laxlabel}{\mathrm{lx}}
\newcommand{\oplaxlabel}{\mathrm{oplx}}
\newcommand{\oplax}{_{\oplaxlabel}}
\newcommand{\lax}{_{\laxlabel}}
\newcommand{\pseudo}{_{\pseudolabel}}
\newcommand{\strict}{_{\strictlabel}}
\newcommand{\arb}{_{\arblabel}}

\newcommand{\fin}{_{\mathrm{fin}}}

\newcommand{\diag}{\mathrm{D}}

\newcommand{\smallsum}{\textstyle{\sum}}
\newcommand{\smallprod}{\textstyle{\prod}}
\usepackage{mdframed}
\usepackage{xcolor}
\usepackage{tcolorbox}

\definecolor{cornellred}{RGB}{196,18,48}
\definecolor{oxfordblue}{RGB}{0,33,71}
\definecolor{cisorange}{RGB}{247,147,33}
\definecolor{purple}{RGB}{96, 24, 168}

\newcommand{\CBPV}{CBPV}
\newcommand{\Vli}[1][\catV]{locally ${#1}$-indexed}
\newcommand{\li}{locally indexed}
\newcommand{\monML}{$\uplambda_{\mathrm{ml}}$}

\newcommand{\toptop}{\top\top} % for discussing Shin-ya's work, in case we want to distinguish

\renewcommand{\vert}{\ast}
\newcommand{\Id}[1]{\mathrm{id}_{#1}}

\newcommand{\shriek}{{}_{\bang}}
\newcommand{\precomp}{{}^*}

\newcommand{\terminal}{1}
\DeclareMathOperator{\bangop}{{!}}
\newcommand{\bang}{\bangop}
\newcommand{\tup}[1]{\langle #1 \rangle }

\newcommand{\catA}{\cat{A}}
\newcommand{\catB}{\cat{B}}
\newcommand{\catC}{\cat{C}}
\newcommand{\catD}{\cat{D}}
\newcommand{\catE}{\cat{E}}

\newcommand{\catM}{\cat{M}}
\newcommand{\catP}{\cat{P}}
\newcommand{\catV}{\cat{V}}
\newcommand{\catW}{\cat{W}}

\newcommand{\mc}[1]{\mathcal{#1}}
\newcommand{\A}{\mc{A}}
\newcommand{\B}{\mc{B}}
\newcommand{\C}{\mc{C}}
\newcommand{\D}{\mc{D}}
\newcommand{\E}{\mc{E}}

\newcommand{\G}{\mc{G}}
\newcommand{\lito}[1]{\xto[#1]{}}

\DeclareMathOperator{\reindop}{\triangleleft}
\newcommand{\reind}{\reindop} % for reindexing

\newcommand{\opGr}[1]{\mathsf{opGr}\, #1}
\newcommand{\self}[1]{\mathsf{self}\, #1}
\newcommand{\selfname}{\mathsf{self}}

\newcommand{\enr}[1]{\overline{#1}} % the enriched part of a map in LInd

\newcommand{\indI}{I}

\newcommand{\oa}{a}
\newcommand{\ob}{b}
\newcommand{\oc}{c}
\newcommand{\od}{d}
\renewcommand{\oe}{e}

\newcommand{\oA}{A}
\newcommand{\oB}{B}
\newcommand{\oC}{C}
\newcommand{\oD}{D}
\newcommand{\oE}{E}

% ...

% ...

% ...

\newcommand{\oR}{R}

% ...
\newcommand{\oV}{V}

\newcommand{\oX}{X}
\newcommand{\oY}{Y}

\newcommand{\List}{\mathit{L}}

\newcommand{\Exc}{\mathrm{Ex}}

\newcommand{\monS}{S}
\newcommand{\monT}{T}
\newcommand{\st}{t}

\newcommand{\mf}{f}
\newcommand{\mg}{g}
\newcommand{\mh}{h}
\newcommand{\mm}{m}
\newcommand{\mn}{n}
\newcommand{\mj}{j}
\newcommand{\mk}{k}
\newcommand{\ml}{\ell}
\renewcommand{\mp}{p}
\newcommand{\mq}{q}
\newcommand{\mr}{r}
\newcommand{\mt}{t}
\newcommand{\mapu}{u}
\newcommand{\mv}{v}

\newcommand{\funF}{F}
\newcommand{\funG}{G}
\newcommand{\funH}{H}
% ...
\newcommand{\funJ}{J}
\newcommand{\funK}{K}
\newcommand{\funL}{L}
\newcommand{\funP}{P}

\newcommand{\funR}{R}
% ...
\newcommand{\funU}{U}

\newcommand{\funf}{f}
\newcommand{\fung}{g}
\newcommand{\funh}{h}

\newcommand{\Sub}{\mathrm{Sub}}
\newcommand{\cod}{\mathrm{cod}}
\newcommand{\arr}{{}^{{\to}}}
\newcommand{\lift}[1]{\dt{#1}}
\newcommand{\oTotal}{E}
\newcommand{\oBase}{B}
\newcommand{\fP}{P}
\newcommand{\fQ}{Q}
\newcommand{\fp}{p}
\newcommand{\fq}{q}

\newcommand{\horseshoe}{\supset}
\newcommand{\cbvcat}{\mathbf{\uplambda ML}}

\newcommand{\cbpvcat}{\mathbf{CBPV}}

\newcommand{\EMModel}[2]{\mathcal{E\!M}(#2)}

\newcommand{\states}{S}

\newcommand{\cartesian}[1]{(#1, {\times}, \terminal)}

\newcommand{\biccc}[1]{(#1, {\times}, \terminal, {\To}, 0, {+})}

\newcommand{\Syn}{\mathbf{Syn}}
\newcommand{\FOSyn}{\mathbf{FOSyn}}

\newcommand{\sig}{\mathcal{S}}

\renewcommand{\v}{\mathsf{v}}
\renewcommand{\c}{\mathsf{c}}
\renewcommand{\k}{\mathsf{k}}

\newcommand{\binddot}{\, . \,}

% ...

\newcommand{\tK}{K}
\newcommand{\tM}{M}
\newcommand{\tN}{N}
% ...
\newcommand{\tV}{V}

\newcommand{\typeA}{A}
\newcommand{\typeB}{B}

\newcommand{\compT}[1]{\overline{#1}}